\documentclass[showpacs,twocolumn,pra]{revtex4}

\usepackage{epsfig}
\usepackage{amsmath}

\usepackage{amsfonts}
\usepackage{amssymb}

\begin{document}

%
%
%

%

\title{Gaussian sharp-edge diffraction: a  paraxial revisitation of Miyamoto-Wolf's  theory\footnote{To Emil Wolf (1922 - 2018), in \emph{Memoriam}}}

\author{Riccardo Borghi}

\affiliation{Dipartimento di Ingegneria, Universit\`a degli Studi ``Roma Tre''\\
Corresponding author: riccardo.borghi@uniroma3.it}





\begin{abstract}
A ``genuinely'' paraxial version of Miyamoto-Wolf's theory aimed at dealing with 
sharp-edge diffraction under Gaussian beam illumination is presented.
The theoretical analysis is carried out in such a way the well known Young-Maggi-Rubinowicz 
boundary diffraction wave theory can be extended to deal with Gaussian beams in an apparently
straightforward way. The key for achieving such an extension is the introduction of suitable 
``complex angles'' within the integral representations of the geometrical and BDW components of the 
total diffracted wavefield. Surprisingly enough, such a simple (although not rigorously justified)
mathematical generalization seems to work well 
within the complex Gaussian realm. The resulting integrals provide meaningful quantities 
that, once suitably combined, give rise to predictions which are in perfect agreement with 
results already obtained in the past. 
An interesting and still open theoretical question about how to evaluate ``Gaussian geometrical shadows''
for arbitrarily shaped apertures is also discussed.
\end{abstract}


\maketitle

\section{Introduction}
\label{Sec:Introduction}

Emil Wolf is universally considered  the father of classical coherence theory.
He  loved  telling to his guests (I was not an exception 
during a lunch at Rochester's Institute of Optics) the following nice story~\cite{Stroud}:
\begin{quotation}
{\em
In 1956 Born was already in retirement and I was on a visiting appointment at New York
University, still working on our book~\cite{Born/Wolf/1999}. One day I received a letter from Born in which he
asked me why the manuscript was not yet finished. I wrote back saying that the manuscript
is almost completed, except for a chapter on partial coherence on which I was still working.
Born replied at once saying, ``Wolf, who apart from you is interested in coherence? Leave
the chapter out and send the manuscript to the printers.''
\\
I finished the chapter anyway and our book was published in 1959, only a few months
before the invention of the laser, and many of the reviews of our book which were then
appearing stressed that ``Principles of Optics'' contained an account of coherence theory, which
had become of crucial importance to the understanding of some features of laser light.
}
\end{quotation}
The careers of several young scientists  who started studying optics 
in the early nineties  have considerably been influenced 
by the big deal of work Wolf and his co-workers produced over almost four 
decades about optical coherence~\cite{Mandel/Wolf/1995}.
It would then be natural to celebrate Wolf's memory and legacy 
by speaking about classical coherence theory.
However, in the present work I wish  pursuing a maybe unusual different route, which touches a 
different area of optics Wolf gave  fundamental
contributions between the sixties and seventies: the so-called boundary diffracted wave 
(BDW henceforth) theory~\cite[Ch.~8]{Born/Wolf/1999}.

The origin of BDW theory  can be traced back to 1802, when Thomas Young first suggested the 
idea that the boundary 
of an illuminated aperture should act as a secondary light source emitting waves in all 
directions~\cite{Young/1802,Rubinowicz/1957}.    
According to Young's picture, diffraction could be thought of as arising from the superposition
of the field produced by clipping the incoming wave through the laws of geometrical optics and
of the so-called boundary diffraction wave (BDW) which originates from the aperture edge.  
Unfortunately, at that time fYoung's ideas were not adequately supported from a mathematical point 
of view. Fresnel's theory, which  mathematically implements Huygens' superposition principle 
under paraxial approximation,  prevailed. It was only between the end of nineteenth and the beginning of
the twentieth century, that {Maggi}~\cite{Maggi/1888} 
and, independently, { Rubinowicz}~\cite{Rubinowicz/1917}, gave Young's idea the
{mathematical basis} needed for it to rise to the status of a quantitative theory.

The Young-Maggi-Rubinowicz theory concerns only with plane- or spherical-wave illuminations,
for which the decomposition of the diffracted  wavefield into a ``geometrical'' plus a BDW
component can be obtained in a fairly simple way~\cite{Asvestas/1985,Forbes/Asatryan/1996,Gordon/Bilow/2002}. On further invoking  paraxial propagation,
the derivation of the BDW decomposition can be achieved starting  from the two-dimensional Fresnel 
in a way that almost resembles an  academic exercise, as first shown by Hannay~\cite{Hannay/2000}.
Quite recently~\cite{Borghi/2015,Borghi/2016,Borghi/2017}, Hannay's  formulation 
has been used as an effective starting point for revisiting Fresnel's diffraction theory within the 
fairly new theoretical framework of the 
so-called catastrophe optics~\cite{Berry/Upstill/1980,Nye/1999}.

The need of extending BDW theory to deal with impinging wavefields more general than plane or spherical
waves was pointed out by Wolf in a celebrated 1960 paper, coauthored with Kerno Miyamoto, where it is
said~\cite{Miyamoto/Wolf/1962} 
\begin{quotation}
{\em
The  researches of Maggi and Rubinowicz showed conclusively the basic correctness of Young's ideas.
However, their analyses were restricted to cases when the wave incident upon the aperture is plane or 
spherical. Attempts to generalize these results to more general fields have so far not been very successful 
and doubts have in fact been expressed about the possibility of such a  generalization.
As a physical model for diffraction, the Young-Maggi-Rubinowicz theory is intrinsically simple and physically 
appealing. It  relates diffraction directly to the true cause of its origin, namely, the presence of the boundary 
of a diffracting  body. It seems hard to believe that no proper generalization to more complicated fields
exists.    
}
\end{quotation}

Among the  main motivations for extending the Young-Maggi-Rubinowicz theory,  still exposed 
in~\cite{Miyamoto/Wolf/1962}, it is  found that
\begin{quotation}
{\em
The possibility of such a generalization is not
of academic interest alone as the following remarks will
indicate. It is well known that the image of a small
source formed by an optical system has as a rule a
complicated structure. In consequence, calculations of
light distribution in such an image are very laborious.

[...]

In any case such a generalization would give a new
insight into the physical process of image formation.
}
\end{quotation}

In~\cite{Miyamoto/Wolf/1962} Miyamoto and Wolf developed their extension of  the
BDW theory to arbitrary impinging wavefields. In particular, they showed that, on working within
the Kirchhoff diffraction theory, the wavefield diffracted by an aperture in an opaque
plane can be decomposed, under very general conditions, as the superposition of 
(i) a disturbance  originating {\em at the} aperture boundary and (ii) a disturbance expressed through 
{\em the sum} of several contributions originated from geometrical singularities suitably located  
\emph{within} the aperture.
When the illumination is plane or spherical, the number of singularities in the disturbance (ii) reduces to
one or zero, depending on the position of the observation point within the geometrical shadow.
In this way, the results of the Young-Maggi-Rubinowicz theory is then  reproduced.
Despite its formal beauty, Miyamoto-Wolf's theory is not easy to be grasped, especially for readers 
not equipped with adequate mathematical backgrounds.
Also its application to apparently simple incoming disturbances, like for instance Gaussian beams, is far from 
being  trivial, as it was pointed out by some works published between the seventies and 
the eighties~\cite{Otis/1974,Otis/Lachambre/Lit/Lavigne/1977,Takenaka/Kakeya/Fukumitsu/1980,Takenaka/Fukumitsu/1982}.
A possible route to develop a more manageable theory is to invoke paraxial approximation from the beginning.
This has already been done in~\cite{Hannay/2000,Borghi/2015,Borghi/2016,Borghi/2017}  as far as the 
Young-Maggi-Rubinowicz theory is concerned, for which the mathematical formulation
 considerably simplifies within  paraxial approximation. 

In the present paper a ``genuinely paraxial'' version of the Miyamoto-Wolf theory will be developed  
for a Gaussian beam impinging on arbitrarily shaped sharp-edge planar apertures.
In particular, on using the well known representation of Gaussian beams in terms of  complex point sources,
a mathematically simple and geometrically sound theoretical treatment of the scalar diffraction problem 
within paraxial approximation can be derived in an apparently almost straightforward way. However,
such apparent simplicity hides some mathematical subtlies which must be  investigated before
the application to practical cases. This is exactly the aim of the present work, which is structured 
as follows: in Sec.~\ref{Sec:Hannay} a brief {\em r\'esum\'e} of the 
paraxial version of the Young-Maggi-Rubinowicz BDW theory developed according to the prescriptions
given in~\cite{Hannay/2000,Borghi/2015,Borghi/2016,Borghi/2017} is  given.
This will help readers to familiarize with the geometry of the problem and the main notations used 
throughout the paper. At the same time, it will make the paper reasonably self contained. 
The general paraxial Gaussian diffraction theory is then carried out in Sec.~\ref{Sec:Gaussian},
where the main analytical results of the work are also presented. In the same section the principal 
mathematical subtleties of the theory are investigated in the light of some conclusions given by Otis 
in~\cite{Otis/1974}. 
In particular, his analysis of the so-called geometrical wavefield  (which is the field ascribed to the singularities present at the diffraction 
aperture) leaved some unanswered questions which could find an explanation, although not definitive, through the 
approach carried out in the present paper. The practical implementation of the BDW integrals derived in Sec.~\ref{Sec:Gaussian}
is described in Sec.~\ref{Sec:NumericalResults} for a single but significant case, namely the diffraction of Gaussian beams 
by misaligned  and tilted circular apertures. 
Some conclusive words are finally given in Sec.~\ref{Sec:Conclusions} to illustrate the main open questions (of both theoretical and practical
nature) as well as the potential applications of the proposed theoretical approach.


\section{Preliminaries: Hannay's theory for plane and spherical waves}
\label{Sec:Hannay}


First of all it is worth  recalling  Hannay's formulation of paraxial BDW theory 
under plane wave illumination following the formulation and notations used in~\cite{Borghi/2015,Borghi/2016}.
\begin{figure}[!ht]
\centerline{\includegraphics[width=5.5cm,angle=-90]{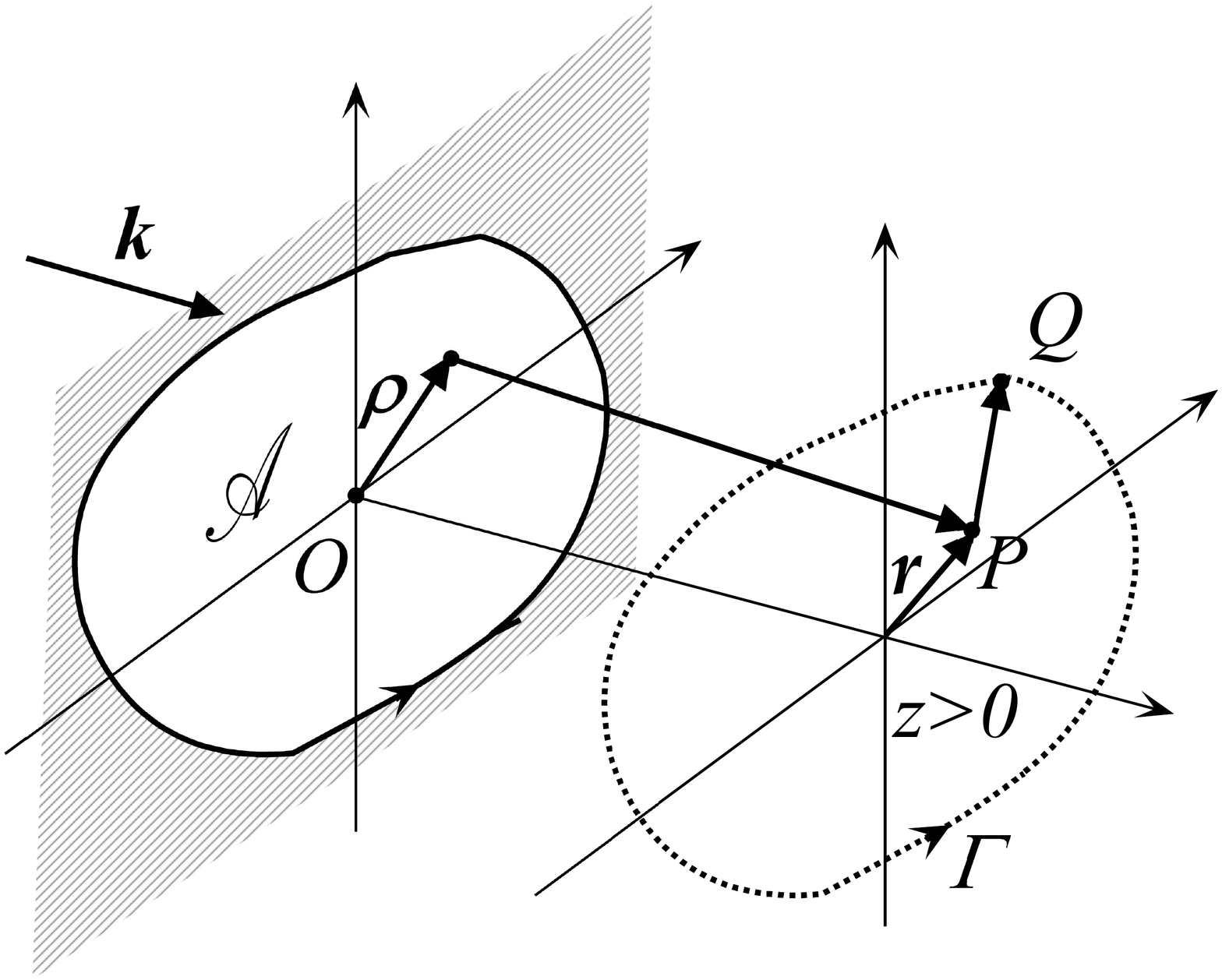}}
\caption{The geometry for the Fresnel integral evaluation.}
\label{Fig:Fresnel.1}
\end{figure}

The geometry of the problem is skecthed in Fig.~\ref{Fig:Fresnel.1}: 
a monochromatic plane wave with wavenumber $k$ orthogonally impinges  on an opaque 
transverse plane with a sharp-edge aperture $\mathcal{A}$.
A unitary (in suitable units) amplitude of the 
incident plane wave will be assumed. The disturbance at a distance $z>0$ from the aperture
is, within the cylindrical reference frame $(\boldsymbol{r};z)$, given by the Fresnel integral
\begin{equation}
\label{Eq:FresnelPropagatorConvolution.1}
\psi(\boldsymbol{r};z)\,=\,
-\frac{\mathrm{i}\,k}{2\pi\,z}\,
\int_{\boldsymbol{\rho}\in\mathcal{A}}\,
\mathrm{d}^2\rho\,
\exp\left[\frac{\mathrm{i}k}{2z}\,(\boldsymbol{r}-\boldsymbol{\rho})^2\right]\,,
\end{equation}
where a factor $\exp(\mathrm{i} kz)$ has been omitted for simplicity.
The two-dimensional (2D henceforth) integral into 
Eq.~(\ref{Eq:FresnelPropagatorConvolution.1}) can be transformed into a one-dimensional
(1D henceforth) contour integral 
%
simply on making the variable change 
$\boldsymbol{\rho}\,\to\,\boldsymbol{R}=\boldsymbol{\rho}\,-\,\boldsymbol{r}$, so that
%
\begin{equation}
\label{Eq:FresnelPropagatorConvolution.1.New}
\psi(\boldsymbol{r};z)\,=\,
-\frac{\mathrm{i}\,k}{2\pi z}\,
\int_{\boldsymbol{\rho}\in\mathcal{A}}\,
\mathrm{d}^2R\,
\exp\left(\frac{\mathrm{i}k}{2z}\,R^2\right)\,.
\end{equation}
\begin{figure}[!ht]
\centerline{\includegraphics[width=3.5cm,angle=-90]{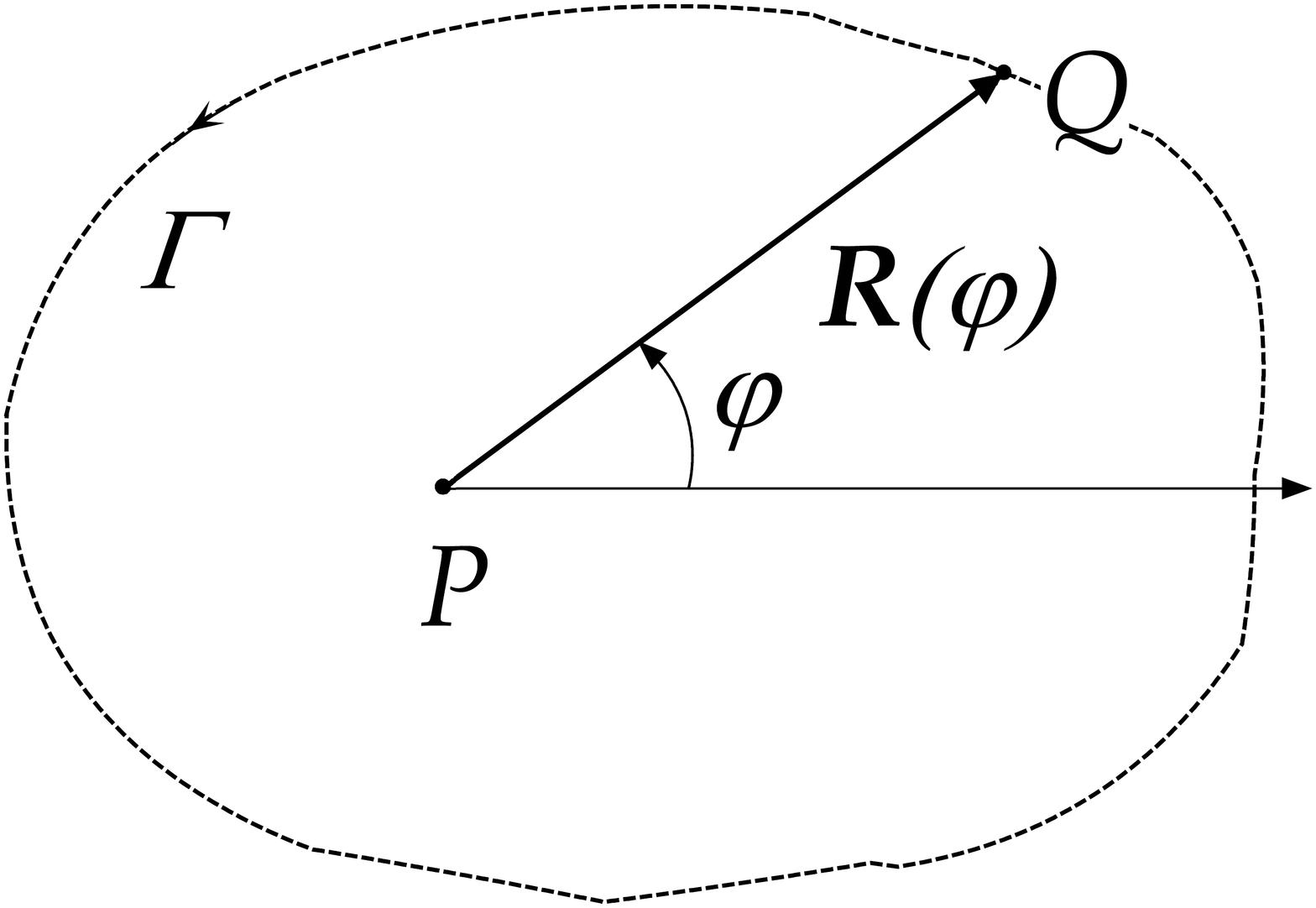}}
\caption{Polar reference frame for the evaluation of paraxial integral in 
Eqs.~(\ref{Eq:FresnelPropagatorConvolution.3.New.2}) and~(\ref{Eq:FresnelPropagatorConvolution.3.New.3}).}
\label{Fig:BDW.2}
\end{figure}

On introducing the polar reference frame  $(R,\varphi)$ shown in 
Fig.~\ref{Fig:BDW.2}, 
Eq.~(\ref{Eq:FresnelPropagatorConvolution.1.New}) then becomes 
\begin{equation}
\label{Eq:FresnelPropagatorConvolution.2.New}
\begin{array}{l}
\displaystyle
\psi(\boldsymbol{r};z)\,=\,
-\frac{\mathrm{i}\,k}{2\pi z}\,
\oint_{\Gamma}\,
\mathrm{d}\varphi\,
\int_0^{R(\varphi)}\,
R\,\mathrm{d}R\,
\exp\left(\frac{\mathrm{i}k}{2z}\,R^2\right)\,=\,\\
\\
\displaystyle
\,=\,
\frac{1}{2\pi}\,
\oint_\Gamma\,\mathrm{d}\varphi\,
\left[1\,-\,\exp\left(\frac{\mathrm{i}k}{2z}\,R(\varphi)^2\right)\right]\,,
\end{array}
\end{equation}
where the function $R=R(\varphi)$  is a polar representation of the boundary
$\Gamma$.
Equation~(\ref{Eq:FresnelPropagatorConvolution.2.New}) will now be recast 
in the form of the following identity:
\begin{equation}
\label{Eq:FresnelPropagatorConvolution.3.New.0.1}
\begin{array}{l}
\displaystyle
-\frac{\mathrm{i}\,k}{2\pi\,z}
\int_{\boldsymbol{\rho}\in\mathcal{A}}\,
\mathrm{d}^2\rho\,
\exp\left[\frac{\mathrm{i}k}{2z}\,(\boldsymbol{r}-\boldsymbol{\rho})^2\right]
=
\psi_G(\boldsymbol{r})\,+\,
\psi_{\rm BDW}(\boldsymbol{r};u)\,,
\end{array}
\end{equation}
where 
\begin{equation}
\label{Eq:FresnelPropagatorConvolution.3.New.2}
\begin{array}{l}
\displaystyle
\psi_G(\boldsymbol{r})\,=\,\frac 1{2\pi}\,\oint_{\Gamma}\,\mathrm{d}\varphi\,,
\end{array}
\end{equation}
and
\begin{equation}
\label{Eq:FresnelPropagatorConvolution.3.New.3}
\begin{array}{l}
\displaystyle
\psi_{\rm BDW}(\boldsymbol{r};u)\,=\,-\frac{1}{2\pi}\,\oint_\Gamma\,\mathrm{d}\varphi\,
\exp\left(\frac{\mathrm{i}u}{2}\,R(\varphi)^2\right)\,.
\end{array}
\end{equation}
%
Here the dimensionless parameter $u=k\ell^{2}/z$ will be henceforth identified with the 
Fresnel number, the parameter $\ell$ being a sort of ``natural'' unit length characteristic of the 
aperture $\mathcal{A}$. For instance, if $\mathcal{A}$ were circularly shaped then $\ell$ would
certainly be identified with the aperture radius.

Equations~(\ref{Eq:FresnelPropagatorConvolution.3.New.0.1})  - (\ref{Eq:FresnelPropagatorConvolution.3.New.3})  will play a key role throughout the present paper.
In particular, the function $\psi_G(\boldsymbol{r})$ defined into Eq.~(\ref{Eq:FresnelPropagatorConvolution.3.New.2}) 
is called ``geometrical wavefield'' and coincides with the characteristic function of the aperture $\mathcal{A}$, i.e., 
\begin{equation}
\label{Eq:FresnelPropagatorConvolution.3.New.3.1}
\begin{array}{l}
\displaystyle
\psi_G(\boldsymbol{r})\,=\, 
\left\{
\begin{array}{lr}
1 & \boldsymbol{r} \in \mathcal{A}\,,\\
&\\
0 & \boldsymbol{r} \notin \mathcal{A}\,.
\end{array}
\right.
\end{array}
\end{equation}
In other words, $\psi_G(\boldsymbol{r})$ represents the field produced, 
according to the laws of geometrical optics,  by clipping the incident plane wave by the aperture $\mathcal{A}$.

The wavefield $\psi_{\rm BDW}$ is that generated by the sharp edge $\Gamma$
and represents Young's wavelets  which have to be superimposed to $\psi_G$
to retrieve the total diffracted field.
The application of the plane wave paraxial BDW theory has already produced 
some interesting results. An important connection  with catastrophe optics has 
already been pointed out in~\cite{Borghi/2015,Borghi/2016,Borghi/2017}, 
where it is also shown how to  build up analytical estimates 
of the diffracted field at observation points which are located in the neighborhood 
of the field singularities. The latter are the geometrical boundary shadow and the caustics 
produced at the geometrical  evolute of the projection of the diffracting aperture on the transverse 
observation plane. 
In particular, the asymptotics treatment of the integral~(\ref{Eq:FresnelPropagatorConvolution.3.New.3}) 
promoted in~\cite{Borghi/2016} allowed  unexpected and interesting features of 
suitably heart-shaped apertures to be grasped~\cite{Borghi/2017}. 
Shortly after, such peculiar properties have also been experimentally confirmed~\cite{Wang/Xie/Ye/Sun/Wang/Feng/Han/Kan/Y. Zhang/2017,Ye/Xie/Wang/Feng/Sun/Zhang/2018,Borghi/2018}.

The ``genuinely paraxial''  Gaussian version of Miyamoto-Wolf's theory promised at the beginning of the paper 
will now be carried out as a suitable generalization of the BDW theory described so far.

\section{Gaussian beams diffraction by sharp-edge apertures}
\label{Sec:Gaussian}

\begin{figure}[!ht]
\centerline{\includegraphics[width=6cm,angle=-90]{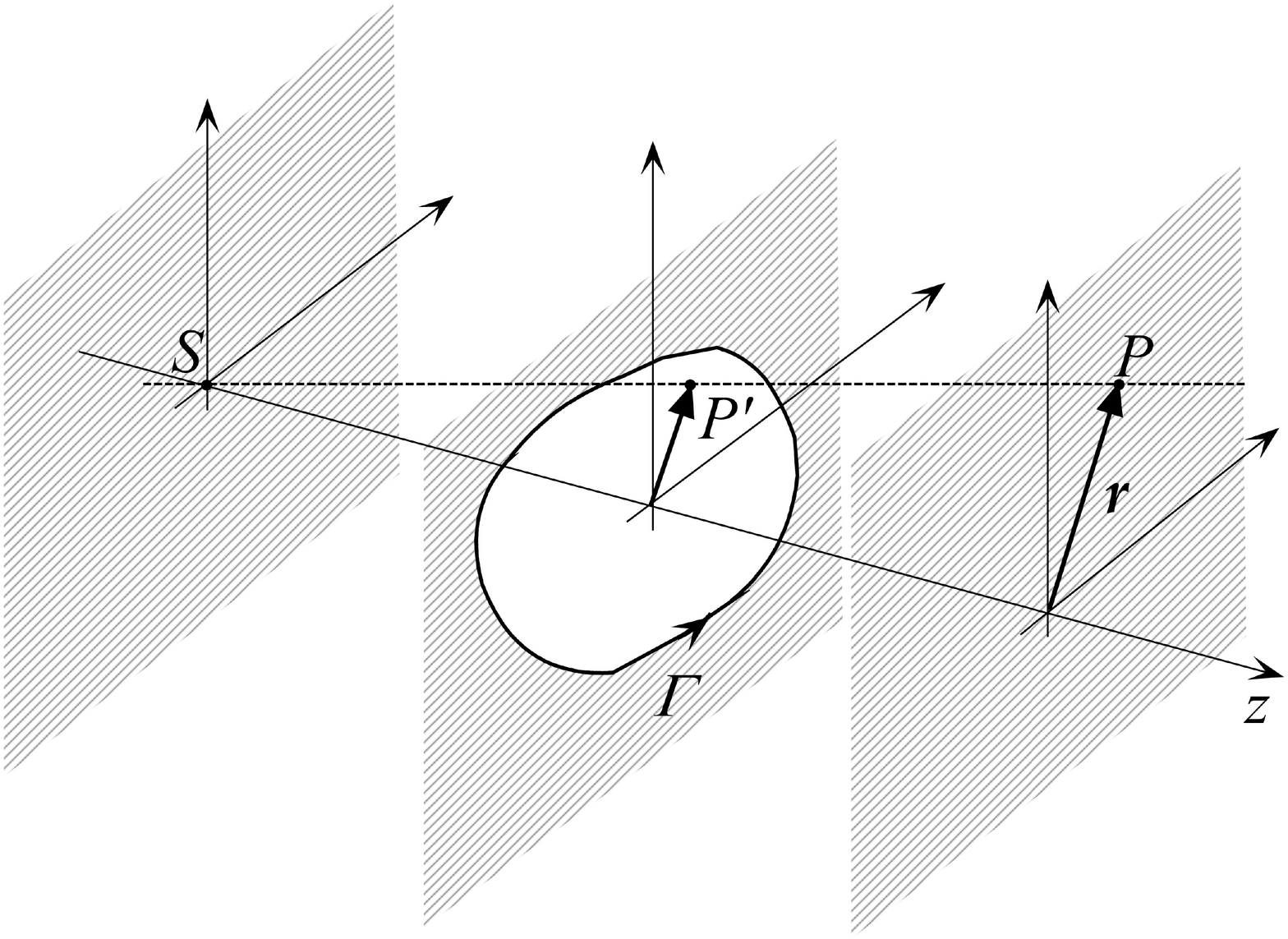}}
\caption{Geometry for point source illumination.}
\label{Fig:PointSource.1}
\end{figure}

The first step is to extend the plane-wave paraxial theory developed in the previous section to 
a spherical wave generated by a point source placed at a distance
$D$ from the aperture plane, as sketched in Fig.~\ref{Fig:PointSource.1}.
Without loss of generality, the $z$-axis will be chosen  to contain the point source. 

In this way the impinging wavefield  across the aperture $\mathcal{A}$  is then
\begin{equation}
\label{Eq:Gaussian.0.1}
\begin{array}{l}
\displaystyle
\psi_i(\boldsymbol{r})\,=\,
\exp\left(\dfrac{\mathrm{i} k}{2D}\,r^2\right)\,,
\end{array}
\end{equation}
where an amplitude factor $1/D$ has been omitted.
It is  straightforward  to prove that the diffracted wavefield 
at the transverse plane $z>0$ is~\cite{Hannay/2000}
\begin{equation} 
\label{Eq:Gaussian.0.2}
\begin{array}{l}
\psi(\boldsymbol{r};z)\,=\,\dfrac{D}{z+D}\,\exp\left(\dfrac{\mathrm{i} k}{2}\,\dfrac{r^2}{z+D}\right)\,\times\,\\
\\
\times
\displaystyle
\left(-\frac{\mathrm{i}\,k}{2\pi}\,\dfrac{z+D}{zD}\right)
\int_{\boldsymbol{\rho}\in\mathcal{A}}\,
\mathrm{d}^2\rho\,
\exp\left[\frac{\mathrm{i}k}{2}\,\dfrac{z+D}{zD}\,\left(\boldsymbol{\rho}\,-\,\dfrac D{z+D}\boldsymbol{r}\right)^2\right]\,,
\end{array}
\end{equation}
where the transverse vector $\boldsymbol{r}_C=\dfrac D{z+D}\boldsymbol{r}$ inside the integral 
defines the position of the point $P'$ corresponding to the intersection between the aperture plane and the line 
connecting $S$ and $P$, as sketched in Fig.~\ref{Fig:PointSource.1}. 

Equation~(\ref{Eq:Gaussian.0.2}) shows that the diffracted wavefield is basically the product of (i) the field the source
$S$ would produce in absence of the aperture $\mathcal{A}$ at a distance $z$ from the aperture plane and (ii) the field a unit-amplitude 
plane wave, orthogonally impinging on $\mathcal{A}$, would produce at the observation point $P'$ of a transverse
plane placed at the distance 
$z_C=(z^{-1}+D^{-1})^{-1}$ from the aperture plane.
On taking Eqs.~(\ref{Eq:FresnelPropagatorConvolution.3.New.0.1}) - (\ref{Eq:FresnelPropagatorConvolution.3.New.3}) 
into account, the diffracted field in Eq.~(\ref{Eq:Gaussian.0.2})
can then be recast as follows:
\begin{equation} 
\label{Eq:Gaussian.0.2.1}
\begin{array}{l}
\psi(\boldsymbol{r};z)\,=\,\dfrac{D}{z+D}\,\exp\left(\dfrac{\mathrm{i} k}{2}\,\dfrac{r^2}{z+D}\right)\,
\displaystyle
\left[
\psi_G(\boldsymbol{r}_C)\,+\,
\psi_{\rm BDW}(\boldsymbol{r}_C;u_C)
\right]\,,
\end{array}
\end{equation}
with 
\begin{equation} 
\label{Eq:Gaussian.0.2.1.1}
\begin{array}{l}
u_C\,=\,\dfrac{k\ell^2}{z_C}\,=\,\dfrac{k\ell^2}{z}\,\left(1+\dfrac zD\right)\,=\,u\,\left(1+\dfrac zD\right)\,,
\end{array}
\end{equation}
denoting the new Fresnel number.

%
%

Equations~(\ref{Eq:Gaussian.0.2.1}) and~(\ref{Eq:Gaussian.0.2.1.1}) represent the key to introduce the paraxial Gaussian BDW theory.
To this end, it must be recalled that 
the wavefield associated to a Gaussian beam formally 
coincides with that produced, within paraxial approximation, by a point source placed somewhere at a ``complex location.''
\begin{figure}[!ht]
\centerline{\includegraphics[width=6cm,angle=-90]{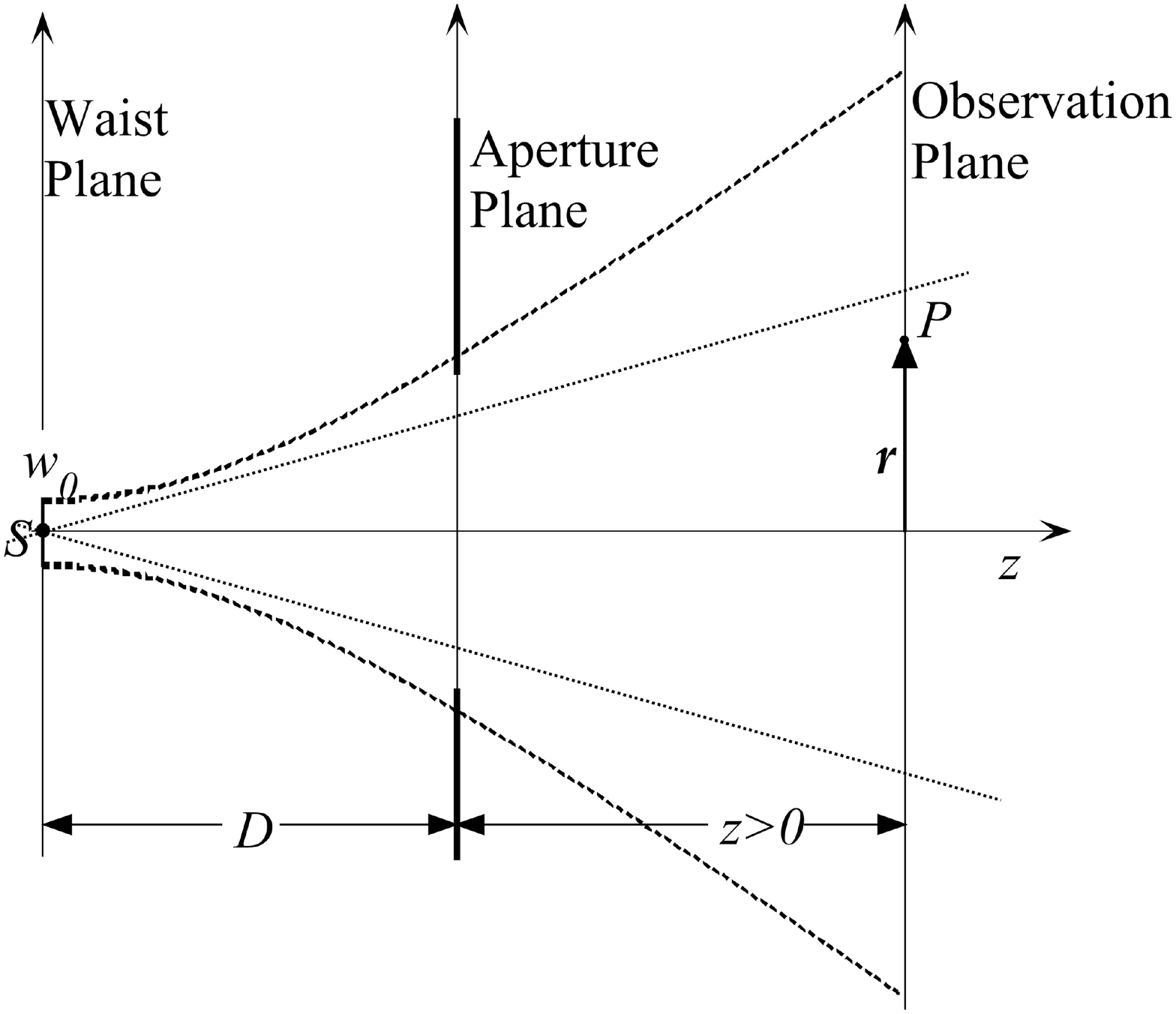}}
\caption{Geometry for Gaussian beam illumination.}
\label{Fig:PointSource.2}
\end{figure}

Consider the geometrical situation depicted in Fig.~\ref{Fig:PointSource.2}:
a Gaussian beam having spot size $w_0$ has its waist plane at a distance $D$ from the aperture plane.
The mean propagation direction of the Gaussian beam does coincide with the $z$-axis of the cylindrical reference frame
$(\boldsymbol{r};z)$. 
For what has been said above, to derive the diffracted wavefield all we have to do is to replace, into Eqs.~(\ref{Eq:Gaussian.0.2.1}) and~(\ref{Eq:Gaussian.0.2.1.1}), 
the {\em real} quantity $D$ by the {\em complex} quantity $D\,-\,\mathrm{i} L$, where $L=kw^2_0/2$,  the so-called Rayleigh length,
will be assumed as the ``natural'' unit to measure all longitudinal distances. Accordingly, on first applying the transformation
$D\to D-\mathrm{i} L$ into Eq.~(\ref{Eq:Gaussian.0.2.1}) and then on formally letting $L=1$, after straightforward 
algebra the Gaussian diffracted wavefield can {\em formally} be expressed as follows:
\begin{equation}
\label{Eq:Gaussian.0.6}
\begin{array}{l}
\psi(\boldsymbol{r};z)\,=\,\dfrac{1+\mathrm{i} D}{1+\mathrm{i} (z+D)}\,\exp\left(-\dfrac{r^2/w_0^2}{1+\mathrm{i} (z+D)}\right)\,\times\,\\
\\
\times
\displaystyle
\,[
\psi_G(\boldsymbol{r}_C)\,+\,
\psi_{\rm BDW}(\boldsymbol{r}_C;u_C)
]\,,
\end{array}
\end{equation}
where both $\boldsymbol{r}_C$ and $u_C$ are now {\em complex } quantities, and precisely
\begin{equation}
\label{Eq:Gaussian.0.6.1}
\left\{
\begin{array}{l}
\boldsymbol{r}_C\,=\,\dfrac{1+\mathrm{i} D}{1+\mathrm{i} (z+D)}\,\boldsymbol{r}\,,\\
\\
u_C\,=\,\dfrac{1+\mathrm{i} (z+D)}{1+\mathrm{i} D}\,u\,.
\end{array}
\right.
\end{equation}
Note that the Fresnel number $u$ can also be expressed in terms of dimensionless quantities as follows (remember that  $L=1$):
%
%
\begin{equation}
\label{Eq:Gaussian.0.6.1.2}
\begin{array}{l}
\dfrac u2\,=\,\dfrac 1z\,\left(\dfrac{\ell}{w}\right)^2\,,
\end{array}
\end{equation}
which appears to be dependent only on the propagation distance measured in terms of the Rayleigh length
and the aperture ``size'' measured in terms of the Gaussian beam spot-size.

Equation~(\ref{Eq:Gaussian.0.6}) is one of the main result of the present paper. It allows Gaussian sharp-edge diffraction to 
be {\em formally } derived from the plane-wave sharp-edge diffraction provided that the functions $\psi_G$ and $\psi_{\rm BDW}$,
which have been defined into Eqs.~(\ref{Eq:FresnelPropagatorConvolution.3.New.2}) and~(\ref{Eq:FresnelPropagatorConvolution.3.New.3})
only for real values of their arguments, can be analytically continued into the complex realm.
To this end, it must be stressed how the geometrical interpretation of Fig.~\ref{Fig:BDW.2} concerning the angular integration variable 
$\varphi$ in Eqs.~(\ref{Eq:FresnelPropagatorConvolution.3.New.2}) and~(\ref{Eq:FresnelPropagatorConvolution.3.New.3}) seems to be 
no longer valid, since the position of the observation point $P$ is now defined by the complex 2D transverse vector of Eq.~(\ref{Eq:Gaussian.0.6.1}).
A possibility to solve this problem is to express the angle $\mathrm{d}\varphi$ through a parametric representation of the aperture boundary $\Gamma$.
On again referring to Fig.~\ref{Fig:BDW.2}, consider the position of a typical point $Q$ on the curve $\Gamma$ to be function of a parameter 
$t$ ranging within a real  interval $\mathcal{I}$. Let $Q=Q(t)$ denotes such parametrization. Then the transverse vector
$\boldsymbol{R}=\overrightarrow{PQ}$ will also be a function of $t$. 
\begin{figure}[!ht]
\centerline{\includegraphics[width=4cm,angle=-90]{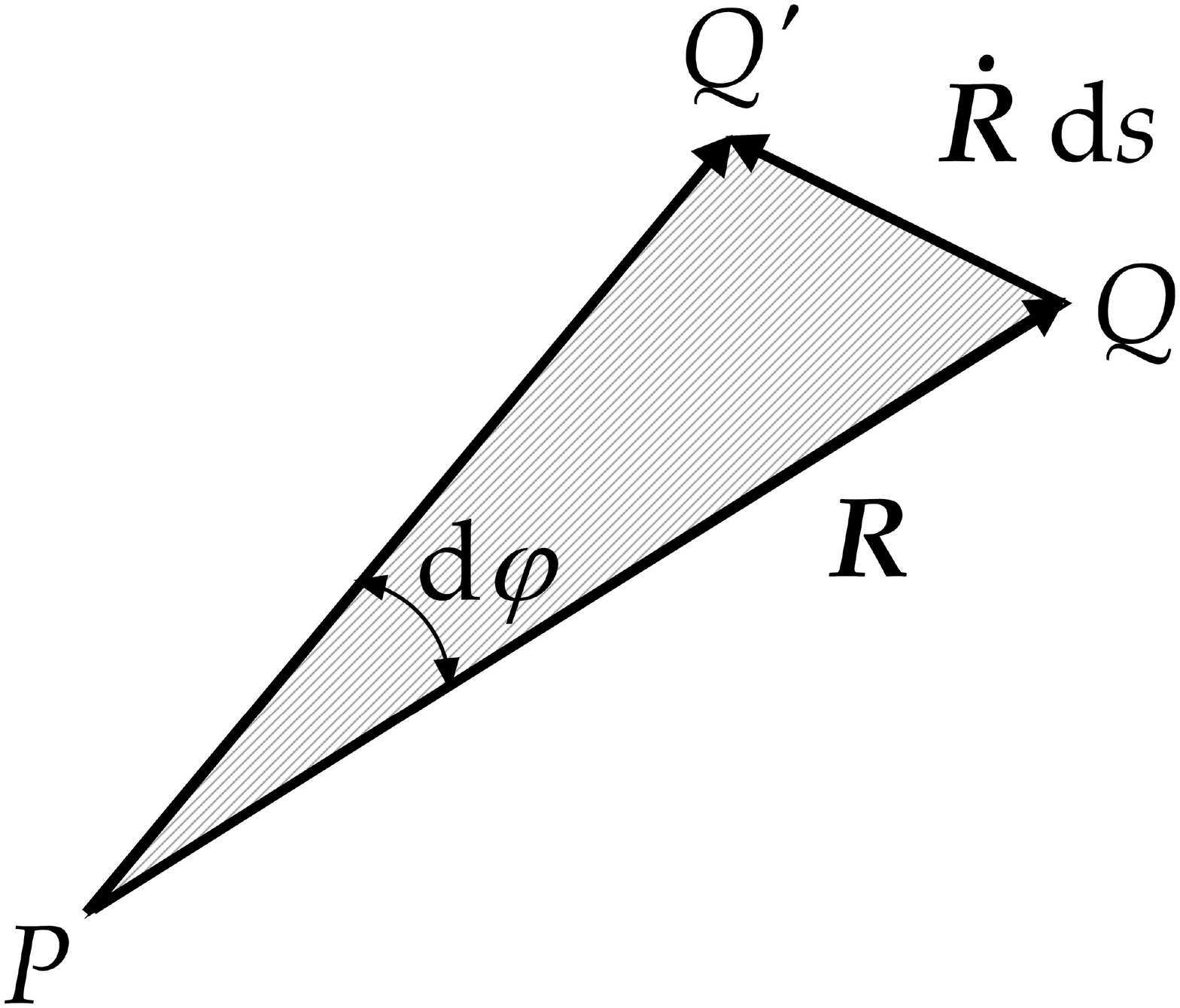}}
\caption{The infinitesimal angle $\mathrm{d}\varphi$ can be evaluated via a parametric representation, say $\boldsymbol{Q}=\boldsymbol{Q}(t)$, 
of the boundary $\Gamma$.}
\label{Fig:Gaussian.0.1}
\end{figure}

On considering two subsequent positions $Q$ and $Q'$ infinitely close in $t$, as sketched in Fig.~\ref{Fig:Gaussian.0.1}, elementary geometry gives at once 
\begin{equation}
\label{Eq:Gaussian.0.6.1.2.1}
\begin{array}{l}
\displaystyle
\mathrm{d}\varphi\,=\,\mathrm{d} t\,
\dfrac{\boldsymbol{R}\times\dot{\boldsymbol{R}}}
{\boldsymbol{R}\cdot\boldsymbol{R}}\,,
\end{array}
\end{equation}
where the dot denotes the derivative with respect to the parameter $t$ and the cross product should be 
intended as the sole $z$-component, being both vectors $\boldsymbol{R}$ and $\dot{\boldsymbol{R}}$ 
purely transverse (i.e., lying on the aperture plane).
Substitution from Eq.~(\ref{Eq:Gaussian.0.6.1.2.1}) into Eqs.~(\ref{Eq:FresnelPropagatorConvolution.3.New.2}) 
and~(\ref{Eq:FresnelPropagatorConvolution.3.New.3}) gives at once
\begin{equation}
\label{Eq:FresnelPropagatorConvolution.3.New.4}
\begin{array}{l}
\displaystyle
\psi_{G}\,=\,\frac{1}{2\pi}\,\oint_\Gamma\,\mathrm{d} t\, 
\dfrac{\boldsymbol{R}\times\dot{\boldsymbol{R}}}
{\boldsymbol{R}\cdot\boldsymbol{R}}\,,
\end{array}
\end{equation}
and
\begin{equation}
\label{Eq:FresnelPropagatorConvolution.3.New.5}
\begin{array}{l}
\displaystyle
\psi_{\rm BDW}\,=\,-\frac{1}{2\pi}\oint_\Gamma\,\mathrm{d} t\, 
\dfrac{\boldsymbol{R}\times\dot{\boldsymbol{R}}}
{\boldsymbol{R}\cdot\boldsymbol{R}}\,
\exp\left(\frac{\mathrm{i}u}{2}\,\boldsymbol{R}\cdot\boldsymbol{R}\right)\,,
\end{array}
\end{equation}
respectively.
Equations~(\ref{Eq:FresnelPropagatorConvolution.3.New.4}) and~(\ref{Eq:FresnelPropagatorConvolution.3.New.5}) could  be evaluated,
in principle, also for those {\em complex } values of $\boldsymbol{r}_C$ and $u_C$ given in Eq.~(\ref{Eq:Gaussian.0.6.1}).
They are the main result of the 
present analysis.

An important check about the validity of Eqs.~(\ref{Eq:FresnelPropagatorConvolution.3.New.4}) and~(\ref{Eq:FresnelPropagatorConvolution.3.New.5})
can be done  in the spherical wave limit, which is reached on letting the Gaussian beam spot size to tend to zero. 
Accordingly, since $L\to 0$, the complex factor into Eq.~(\ref{Eq:Gaussian.0.6.1}) tends to the {\em real} limit
$D/(z+D)$, in agreement with Eqs.~(\ref{Eq:Gaussian.0.2})-(\ref{Eq:Gaussian.0.2.1.1}).
From a mere mathematical point of view, to formally consider complex values of the observation point $P$ into the definition of 
$\boldsymbol{R}$ could allow the quantity $\boldsymbol{R}\cdot\boldsymbol{R}$ to vanish for some real values of $t$.
Accordingly, the integrand inside Eqs.~(\ref{Eq:FresnelPropagatorConvolution.3.New.4}) and~(\ref{Eq:FresnelPropagatorConvolution.3.New.5})
would be singular. 

In a 1974 paper~\cite{Otis/1974},  Otis claimed that a Gaussian beam impinging on a typical sharp-edge aperture $\mathcal{A}$
according to the geometry depicted in Fig.~\ref{Fig:PointSource.2} should produce a well defined geometrical shadow, which is obtained
simply by projecting the aperture boundary $\Gamma$ via a suitable  radially symmetric hyperboloid~\cite[Eq.~(46)]{Otis/1974}. 
In particular, the transverse shape of the boundary shadow at the typical propagation plane $z>0$ should  be a  replica of $\Gamma$ 
 scaled by the following (within our dimensionless units) {\em real} factor~\cite[Eq.~(47)]{Otis/1974}:
\begin{equation}
\label{Eq:Otis.47}
\begin{array}{l}
\sqrt{\dfrac{1+ (z+D)^2}{1+ D^2}}\,.
\end{array}
\end{equation}
It must be noted how Otis' conjecture should then imply the integral in Eq.~(\ref{Eq:FresnelPropagatorConvolution.3.New.4}),
once evaluated at complex transverse vectors $\boldsymbol{r}_C=\boldsymbol{r}\,\exp(\mathrm{i}\varphi)$,
to satisfy the following relation:
\begin{equation}
\label{Eq:FresnelPropagatorConvolution.3.New.3.1}
\begin{array}{l}
\displaystyle
\psi_G(\boldsymbol{r}\,\exp(\mathrm{i}\varphi))\,=\, 
\left\{
\begin{array}{lr}
1 & \boldsymbol{r} \in \mathcal{A}\,,\\
&\\
0 & \boldsymbol{r} \notin \mathcal{A}\,.
\end{array}
\right.
\end{array}
\end{equation}
For circular apertures such conjecture can rigorously be proved. To this aim, consider a circular aperture of 
radius $a$, centred on the $z$-axis. On letting $\ell=a$, a simple parametrization of the aperture boundary $\Gamma$
is $Q(t)=(\cos t,\sin t)$, with $t\in[0,2\pi]$.
Moreover, due to the axial symmetry of the problem, it is expected the geometrical field $\psi_G$ to be a radial function 
of the (complex) normalized quantity $\xi$ defined by
\begin{equation}
\label{Eq:GaussianCircle.0}
\begin{array}{l}
\xi\,=\,\dfrac{1+\mathrm{i} D}{1+\mathrm{i} (z+D)}\,\dfrac{r}{a}\,.
\end{array}
\end{equation}
%
On expressing  
Eq.~(\ref{Eq:FresnelPropagatorConvolution.3.New.4}) through Cartesian coordinates, it is easily found that
\begin{equation}
\label{Eq:GaussianCircle.0.1}
\begin{array}{l}
\displaystyle
\psi_G(\xi)\,=\,\dfrac 1\pi\,\int_0^\pi\,
\mathrm{d} t\,\dfrac{1\,-\,\xi\,\cos t}{1\,+\,\xi^2\,-\,2\xi\,\cos t}\,,
\qquad\qquad \xi\in\mathbb{C}\,,
\end{array}
\end{equation}
and in Appendix~\ref{Sec:AppendixA} it is shown that 
\begin{equation}
\label{Eq:GaussianCircle.1}
\begin{array}{l}
\displaystyle
\dfrac 1\pi\,\int_0^\pi\,
\mathrm{d} t\,\dfrac{1\,-\,\xi\,\cos t}{1\,+\,\xi^2\,-\,2\xi\,\cos t}\,=\,
\left\{
\begin{array}{lr}
1 &  |\xi| < 1\,,\\
&\\
0 &  |\xi| > 1\,,
\end{array}
\right.
\end{array}
\end{equation}
which definitely proves Eq.~(\ref{Eq:FresnelPropagatorConvolution.3.New.3.1})
in the case of circular apertures.

\begin{figure}[!ht]
\centerline{\includegraphics[width=7cm,angle=-0]{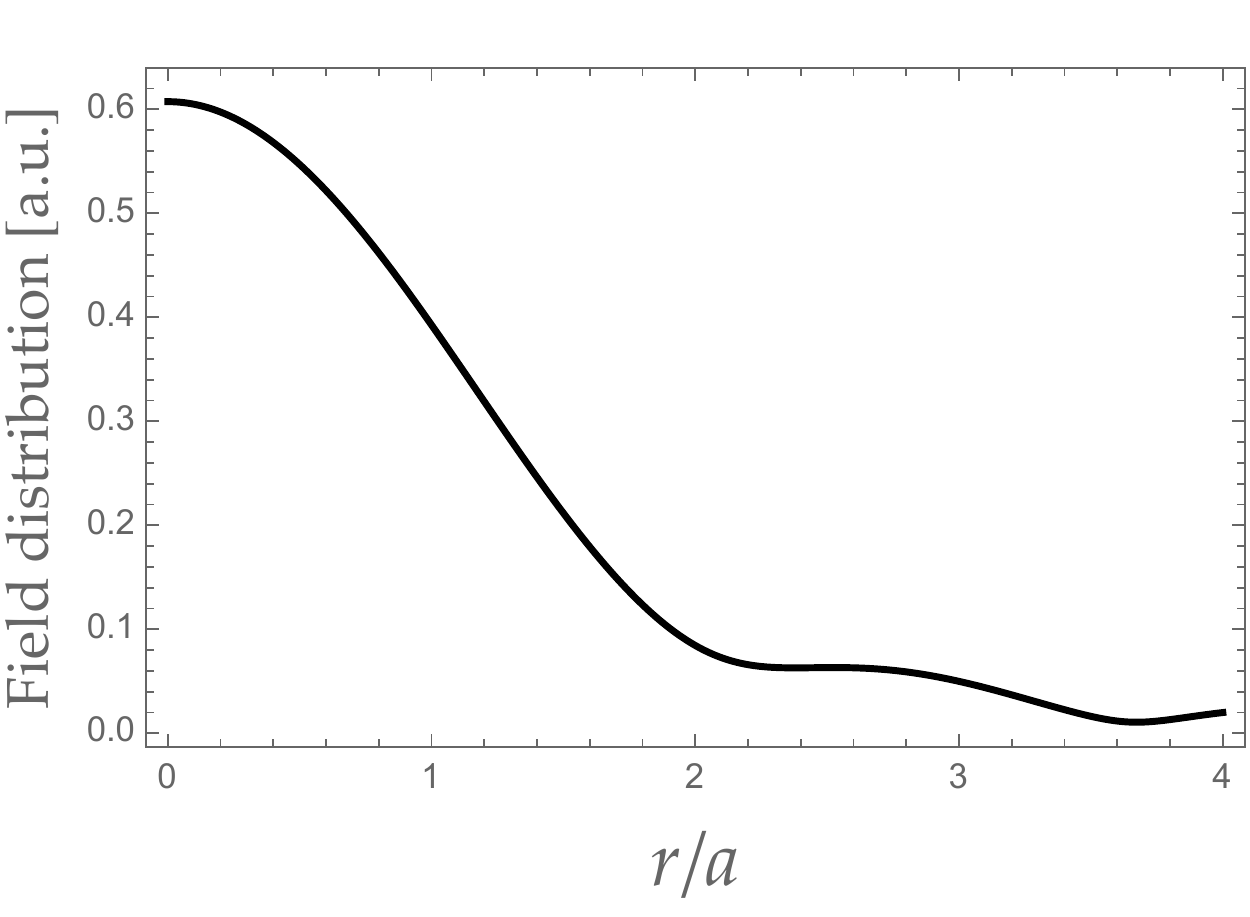}}
\caption{Behaviour of the transverse field distribution of the wavefield produced, at a propagation
distance of one Rayleigh length ($z=1$), via the diffraction of a Gaussian beam 
by a circular aperture placed at the beam waist plane ($D=0$) and having the radius coincident with 
the spot size ($\alpha=1$). The figure
should be compared to Fig.~4 of~\cite{Takenaka/Kakeya/Fukumitsu/1980}.}
\label{Fig:Takenaka}
\end{figure}

As far as the BDW wavefield  is concerned, the substitution of the circle parametrization 
into Eq.~(\ref{Eq:FresnelPropagatorConvolution.3.New.5}) gives at once the integral 
which has already been derived in~\cite{Otis/Lachambre/Lit/Lavigne/1977} on the basis of the 
original Miyamoto-Wolf theory.
To give an idea about the fact that Eqs.~(\ref{Eq:FresnelPropagatorConvolution.3.New.4})
and~(\ref{Eq:FresnelPropagatorConvolution.3.New.5}) provide meaningful quantities when 
applied to the study of the Gaussian diffraction by a circular aperture, in Fig.~\ref{Fig:Takenaka}
it is shown the transverse field distribution of the wavefield produced, at a propagation
distance of one Rayleigh length ($z=1$), via the diffraction of a Gaussian beam 
by a circular aperture placed at the beam waist plane ($D=0$) and having the radius coincident with 
the spot size ($\alpha=1$). An identical situation was considered long ago by Takenaka~\emph{et al.}
in a paper~\cite{Takenaka/Kakeya/Fukumitsu/1980}  where asympotic estimates of the diffracted 
wavefield were  obtained starting from the original Miyamoto-Wolf theory.
In particular, Fig.~\ref{Fig:Takenaka} should be compared to Fig.~4 of~\cite{Takenaka/Kakeya/Fukumitsu/1980}.
The agreement is perfect. 

\begin{figure}[!ht]
\centerline{\includegraphics[width=6cm,angle=-90]{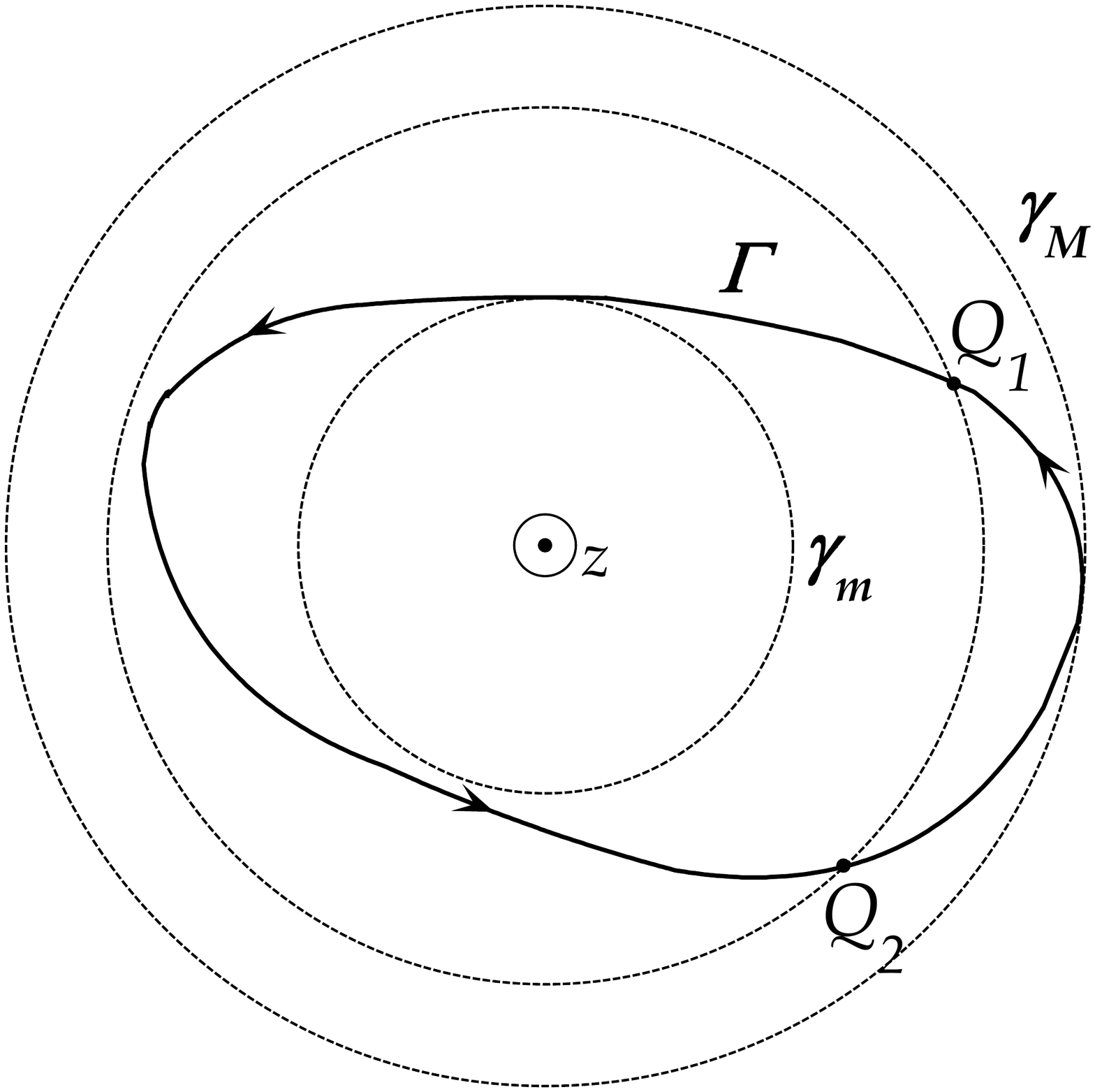}}
\caption{Geometrical wavefield $\psi_G$ for a nonsymmetric aperture.}
\label{Fig:Singularities}
\end{figure}

The results obtained so far would seem to confirm the Otis conjecture about the geometrical intepretation of the
wavefield $\psi_G$  in Eq.~(\ref{Eq:FresnelPropagatorConvolution.3.New.3.1}). 
However, things are considerably more cumbersome as they could appear at first sight. 
Consider a typical aperture $\Gamma$, sketched in Fig.~\ref{Fig:Singularities} and let $Q(t)=[X(t),Y(t)]$
be a suitable parametrization of $\Gamma$. 
We ask 
when the scalar product $\boldsymbol{R}\cdot\boldsymbol{R}$ vanishes on letting the observation point
$P=(\xi,\eta)$   to attain \emph{complex} values of its coordinates $\xi$ and $\eta$ according to Eq.~(\ref{Eq:FresnelPropagatorConvolution.3.New.3.1}). 
To this end, it should be noted that the integrand singularities are the \emph{real} solution of the equation $\boldsymbol{R}\cdot\boldsymbol{R}=0$, i.e.,
\begin{equation}
\label{Eq:GaussianAperture.1}
\begin{array}{l}
\displaystyle
[X(t)\,-\,\xi]^2\,+\,[Y(t)\,-\,\eta]^2\,=\,0\,.
\end{array}
\end{equation}
These solutions can formally be written through the implicit form
\begin{equation}
\label{Eq:GaussianAperture.2}
\begin{array}{l}
\displaystyle
X(t)\,-\,\xi\,=\,\pm\,\mathrm{i}\,[Y(t)\,-\,\eta]\,,
\end{array}
\end{equation}
which immediately  leads to the following \emph{necessary} condition:
\begin{equation}
\label{Eq:GaussianAperture.2}
\begin{array}{l}
\displaystyle
X^2(t)\,+\,Y^2(t)\,=\,|\xi|^2\,+\,|\eta|^2\,.
\end{array}
\end{equation}
Equation~(\ref{Eq:GaussianAperture.2}) has a clear and simple geometrical interpretation, which is depicted in Fig.~\ref{Fig:Singularities}:
circles $\gamma_m$ and $\gamma_M$ are both centred on the $z$-axis (the Gaussian beam mean propagaton direction).
The former is the entirely made by points inside $\Gamma$.
On the contrary, the latter is the smallest circle centred on $z$ which is entirely made by points outside $\Gamma$. For all observation points 
between $\gamma_m$ and $\gamma_M$, Eq.~(\ref{Eq:GaussianAperture.2}) admits at least one real solution
(for example points $Q_1$ and $Q_2$ in Fig.~\ref{Fig:Singularities}).
Figure~\ref{Fig:Singularities} clearly shows what is the main drawback within a general scenario: the mismatch between the axial symmetry
of the incident Gaussian beam and the shape of the diffracting aperture which, apart from the unique case of a coaxial circular hole, cannot 
share the Gaussian axial symmetry at all.
This implies that the geometrical wavefield $\psi_G$ is expected to be identically and rigorously equal to 1, regardless the phase value $\varphi$,
only for the observation points inside the inner circle $\gamma_m$. At the same time, it is expected $\psi_G$ to be 
identically and rigorously equal to 0, regardless the phase value $\varphi$, only for  observation points outside the outer circle $\gamma_M$. 
When the point $P$ is between the two circles, changing the phase $\varphi$
will cause Eq.~(\ref{Eq:GaussianAperture.1}) to be satisfied for some \emph{real} values of $t$, thus making both integrals into 
Eqs.~(\ref{Eq:FresnelPropagatorConvolution.3.New.4}) and~(\ref{Eq:FresnelPropagatorConvolution.3.New.5}) singular.
In other words, the geometrical wavefield is expected to display, for a given $P$, a series of discontinuities on letting $\varphi$ to vary.

A simple example that can be deal with in analytical terms is the replacement of the circular aperture by an \emph{elliptic} aperture 
centred at the Gaussian beam mean direction. Let $\epsilon$ be the ellipse eccentricity, so that $\chi=\sqrt{1+\epsilon^2}>1$ will 
denote the ellipse major half-axis, being still unitary the minor half-axis. For simplicity we shall  consider the evaluation of the geometrical wavefield 
$\psi_G$ at observation points of the form $P=(\xi,0)$, i.e., along the ellipse major axis. Let $\varphi$ the phase of $\xi$.
On using the ellipse parametrization $[X(t),Y(t)]=(\chi\cos t,\sin t)$, with $t\in[0,2\pi]$, the geometrical wavefield turns out to be
\begin{equation}
\label{Eq:GaussianEllipse.1}
\begin{array}{l}
\displaystyle
\psi_G(\xi)\,=\,\dfrac 1{2\pi}\,\int_0^{2\pi}\,
\mathrm{d} t\,\dfrac{\chi\,-\,\xi\,\cos t}
{1\,+\,\xi^2\,-\,2\xi\,\chi\,\cos t\,+\,(\chi^2-1)\,\cos^2t}\,.
\end{array}
\end{equation}
From Equation~(\ref{Eq:GaussianAperture.2}) it follows at once that 
the integrand in Eq.~(\ref{Eq:GaussianEllipse.1}) will be singular if 
\begin{equation}
\label{Eq:GaussianEllipse.1.1}
\begin{array}{l}
\displaystyle
1\,<\, |\xi|\, <\, \chi\,,
\end{array}
\end{equation}
as expected from the above analysis.
Moreover, for a given value of $|\xi|$ it is not difficult to show that the value of the phase $\varphi$ at which the discontinuity occurs can be expressed 
in analytical terms as follows:
\begin{equation}
\label{Eq:GaussianEllipse.1.1.1}
\begin{array}{l}
\displaystyle
\tan\varphi\,=\,\dfrac{\sqrt{\chi^2-|\xi|^2}}{\chi\sqrt{|\xi|^2-1}}\,,\qquad 1\,<\, |\xi|\, <\, \chi\,.
\end{array}
\end{equation}
\begin{figure}[!ht]
\centerline{\includegraphics[width=8cm,angle=-0]{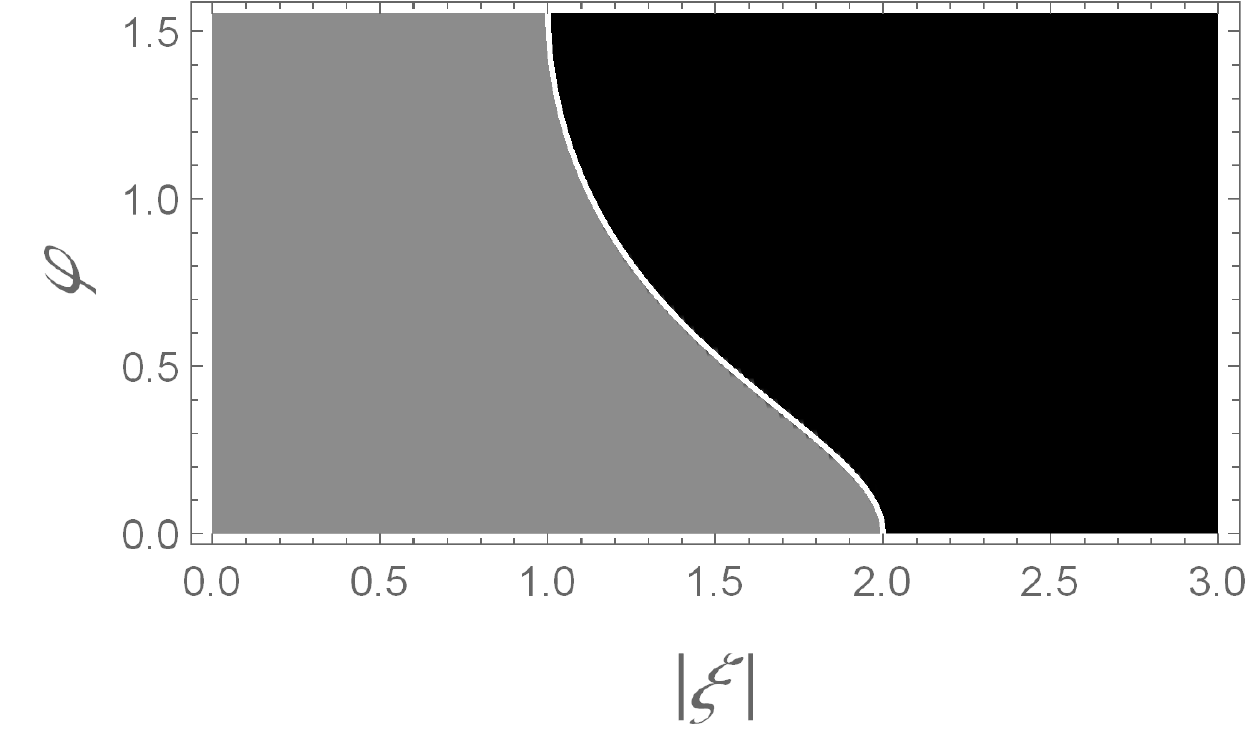}}
\caption{Two-dimensional map of the geometrical wavefield $\psi_G$ for an elliptic aperture with $\chi=2$, 
numerically evaluated via Eq.~(\ref{Eq:FresnelPropagatorConvolution.3.New.4}) at complex observation points $(|\xi|\exp(\mathrm{i}\varphi),0)$.
}
\label{Fig:FigureSingularities}
\end{figure}

\begin{figure}[!ht]
\centerline{\includegraphics[width=8cm,angle=-0]{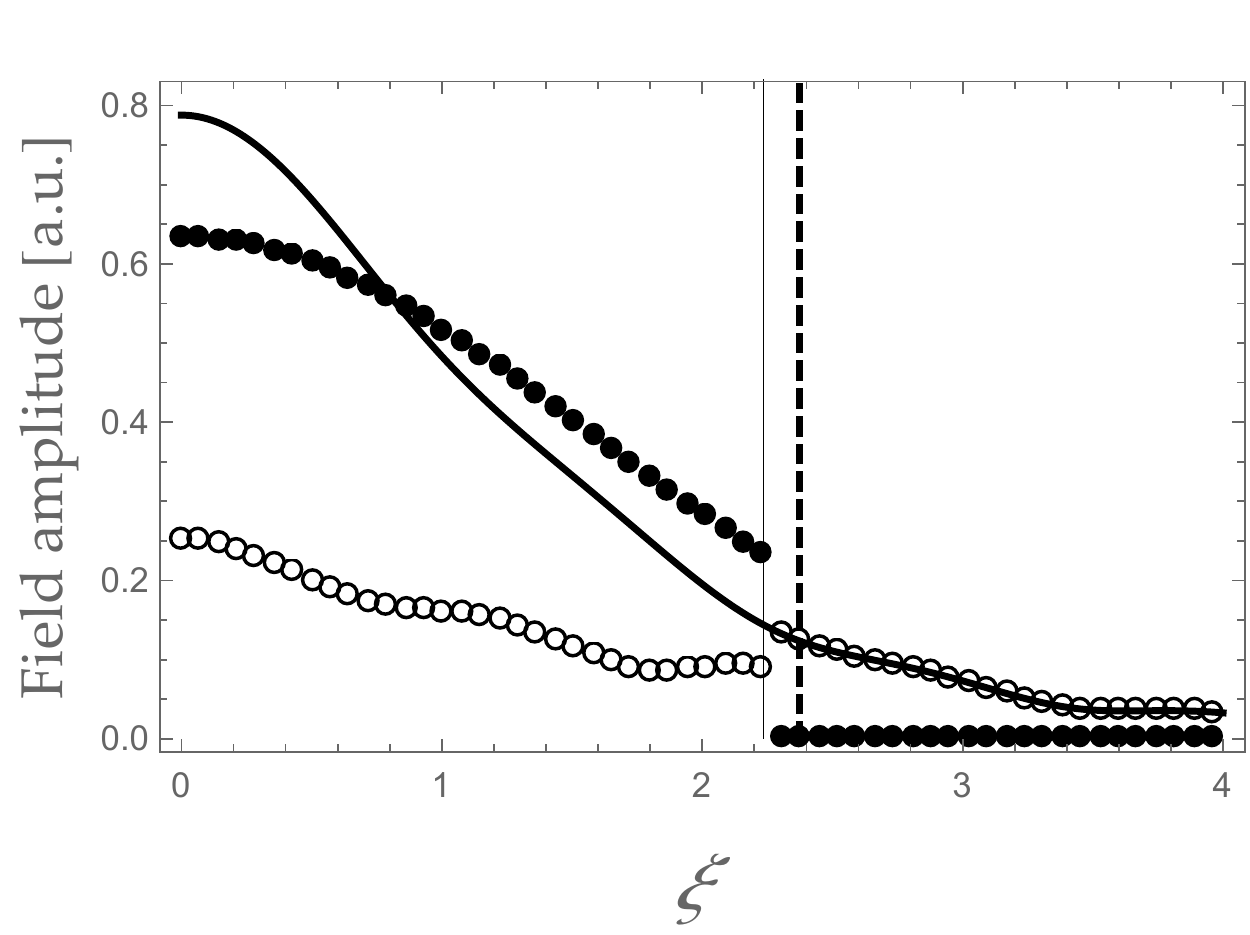}}
\caption{
Behaviours of the field amplitudes of the geometrical (dots), of the BDW (open circles),
and of the total (geometrical plus BDW) diffracted field (solid curve), as functions of the normalized abscissa $\xi$ along 
the major axis of an ellipse with $\chi=3/2$, $\alpha=1$, $z=1$, and $D=3$. 
 The field discontinuity occurs at $\bar\xi\simeq 2.24$. The vertical dashed line represents the Otis 
prediction ($\xi\simeq 2.37$)about the geometrical shadow of the ellipse.}
\label{Fig:EllipseMajorAxis}
\end{figure}

A visual  check of Eq.~(\ref{Eq:GaussianEllipse.1.1.1}) is shown in Fig.~\ref{Fig:FigureSingularities}, where
a two-dimensional map of the geometrical wavefield $\psi_G$ for an elliptic aperture with $\chi=2$, 
numerically evaluated via Eq.~(\ref{Eq:FresnelPropagatorConvolution.3.New.4})
at complex observation points $(|\xi|\exp(\mathrm{i}\varphi),0)$, is shown.
Within the grey region it turns out that $\psi_G=1$, whereas within the black region $\psi_G=0$. The white curve
is just Eq.~(\ref{Eq:GaussianEllipse.1.1.1}).
For the geometrical configuration of Fig.~\ref{Fig:PointSource.2}
 it is trivial to show, from Eq.~(\ref{Eq:GaussianCircle.0}), that the geometrical field jumps must occur at 
values of $\varphi$ given by
\begin{equation}
\label{Eq:GaussianEllipse.1.1.1.1}
\begin{array}{l}
\displaystyle
\tan\varphi\,=\,\dfrac{z}{1+\delta (\delta+z)}\,,
\end{array}
\end{equation}
which, together with Eq.~(\ref{Eq:GaussianEllipse.1.1.1}), gives the position of the geometrical wavefield discontinuity
along the  ellipse major axis, say $\bar\xi$, as follows:
\begin{equation}
\label{Eq:GaussianEllipse.1.1.1.1.1}
\begin{array}{l}
\displaystyle
\bar\xi\,=\,\chi\,\sqrt{\dfrac{1+(z+D)^2}{1+D^2}}\,\sqrt{\dfrac{1+\left(\dfrac z{1+D(z+D)}\right)^2}{1+\chi^2\left(\dfrac z{1+D(z+D)}\right)^2}}
\,.
\end{array}
\end{equation}
To numerically check Eq.~(\ref{Eq:GaussianEllipse.1.1.1.1.1}), in Fig.~\ref{Fig:EllipseMajorAxis} the field amplitudes of the geometrical (dots), the BDW (open circles),
and the total (geometrical plus BDW) diffracted field (solid curve), are shown as functions of the normalized abscissa $\xi$ along 
the major axis of an ellipse with $\chi=3/2$, $alpha=1$, $z=1$, and $D=3$. It can appreciated the discontinuities of both the geometrical and the BDW wavefields
at the value $\bar\xi\simeq 2.24$ theoretically predicted by Eq.~(\ref{Eq:GaussianEllipse.1.1.1.1.1}), whereas the total field turns out to
be continuous. In the same figure the Otis conjecture about the geometrical shadow boundary (occurring at $\xi\simeq 2.37$)
is also shown (vertical dashed line). 

\section{Gaussian beam diffraction from misaligned and titled circular apertures and plates}
\label{Sec:NumericalResults}

\begin{figure}[!ht]
\centerline{\includegraphics[width=6cm,angle=-90]{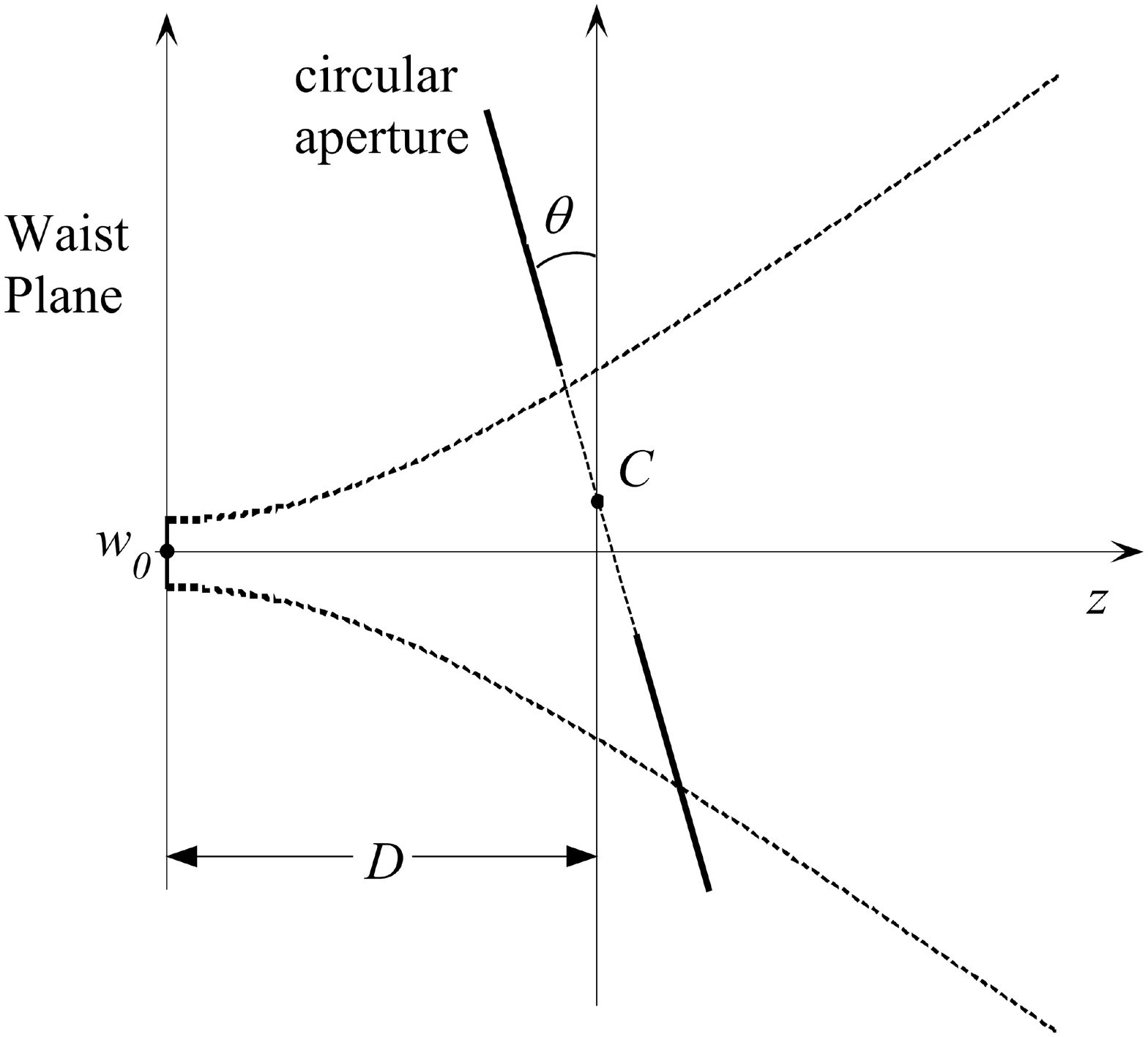}}
\caption{Misalignement of a circular aperture (or plate).}
\label{Fig:Misaligned}
\end{figure}

The present section is devoted to illustrate a  significant practical application of the 
theoretical treatment carried out in the previous sections. 
It is worth starting from some of the beautiful results obtained by Coulson and 
Becknell in~\cite{Coulson/Becknell/1922,Coulson/Becknell/1922b} where, in order 
to experimentally investigate plane-wave diffraction by elliptic opaque plates, suitably 
titled circular plates were then employed.
Such a scenario is depicted in Fig.~\ref{Fig:Misaligned}, where the symbol $\theta$ denotes the 
tilting angle and where the center $C$ of the circular aperture has also been placed off-axis with 
respect to the Gaussian beam mean propagation axis $z$ to account for misalignments. 
For small values of the tilting angle $\theta$, the circular aperture can then approximately be viewed from the
Gaussian beam waist plane as an off-axis elliptic aperture. 
The study of the effects of misalignments of a limiting aperture on the transmitted wavefield
plays a role of pivotal importance, for instance, for  the design of laser communication systems, as it was recently
pointed out in~\cite{Song/Wang/Wu/Ohtsuki/Gurusamy/Kam/2017}. 
It should be stressed again how the theoretical 
approach carried out in the present paper allows, in principle, to deal with arbitrarily
shaped and displaced apertures. To this end, the implementative easiness of Eqs.~(\ref{Eq:FresnelPropagatorConvolution.3.New.4})
and~(\ref{Eq:FresnelPropagatorConvolution.3.New.5}) will first be tested by reproducing some of the results
obtained in a 1969 paper by Pearson~\emph{et al.}, who investigated (theoretically and experimentally)
the paraxial diffraction of a Gaussian beams by a semi-infinite edge~\cite{Pearson/McGill/Kurtin/Yariv/1969}.
\begin{figure}[!ht]
\centerline{\includegraphics[width=5cm,angle=-90]{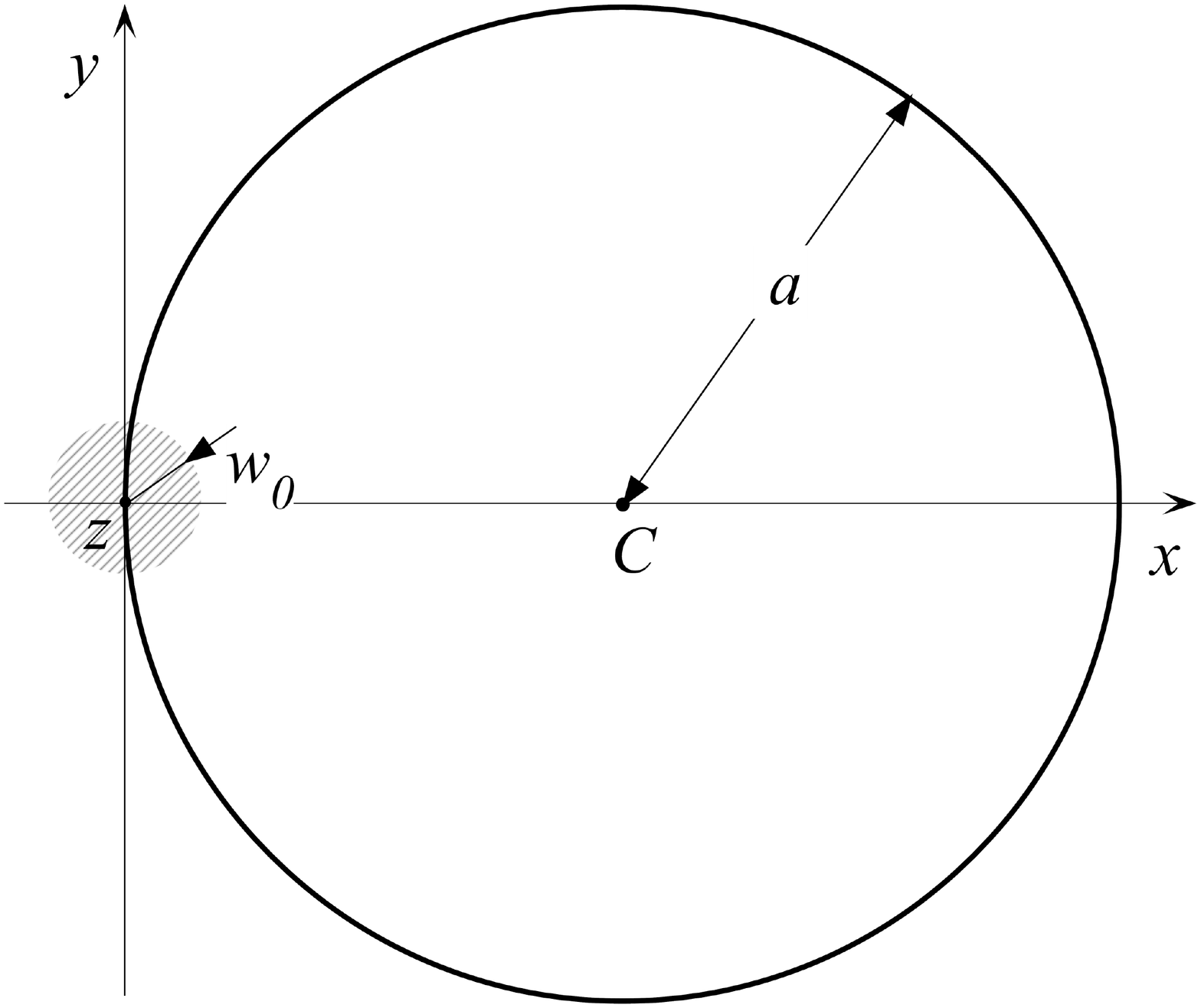}}
\caption{Geometry for reproducing the results of~\cite{Pearson/McGill/Kurtin/Yariv/1969}.}
\label{Fig:YarivSperimentale}
\end{figure}

In Fig.~\ref{Fig:YarivSperimentale} the geometry employed for reproducing the results of~\cite{Pearson/McGill/Kurtin/Yariv/1969} is shown.
The circular aperture is placed in such a way the beam axis is passing through its edge. A transverse Cartesian reference frame $Oxy$
has been introduced with the $x$-axis passing through the aperture centre $C$. The experimental data we have chosen to reproduce are
those shown in Figs.~5 and~6 of~\cite{Pearson/McGill/Kurtin/Yariv/1969}, whose physical parameters are listed in Tab.~\ref{Tab:table.1}.
The beam wavelength is $\lambda=6328$~\AA.
\begin{table}[!ht]
\centerline{
    \begin{tabular}{c|c|c}
    \hline
Parameter    & Fig.~5 & Fig.~6  \\ \hline \hline
	$L$ 		&	 776 (cm)		&	96 (cm) \\ \hline 
	$D$ 		&	 1480 (cm)		&	 101 (cm) \\ \hline 
	$z$ 		&	 100 (cm)		&	 100 (cm) \\ \hline 
    \hline
    \end{tabular}
}
\caption{Physical parameters used to produce the results in Figs.~5 and~6 of~\cite{Pearson/McGill/Kurtin/Yariv/1969}.
The wavelength is $\lambda=6328$~\AA}
\label{Tab:table.1}
\end{table}
\begin{figure}[!ht]
\centerline{\includegraphics[width=8cm,angle=-0]{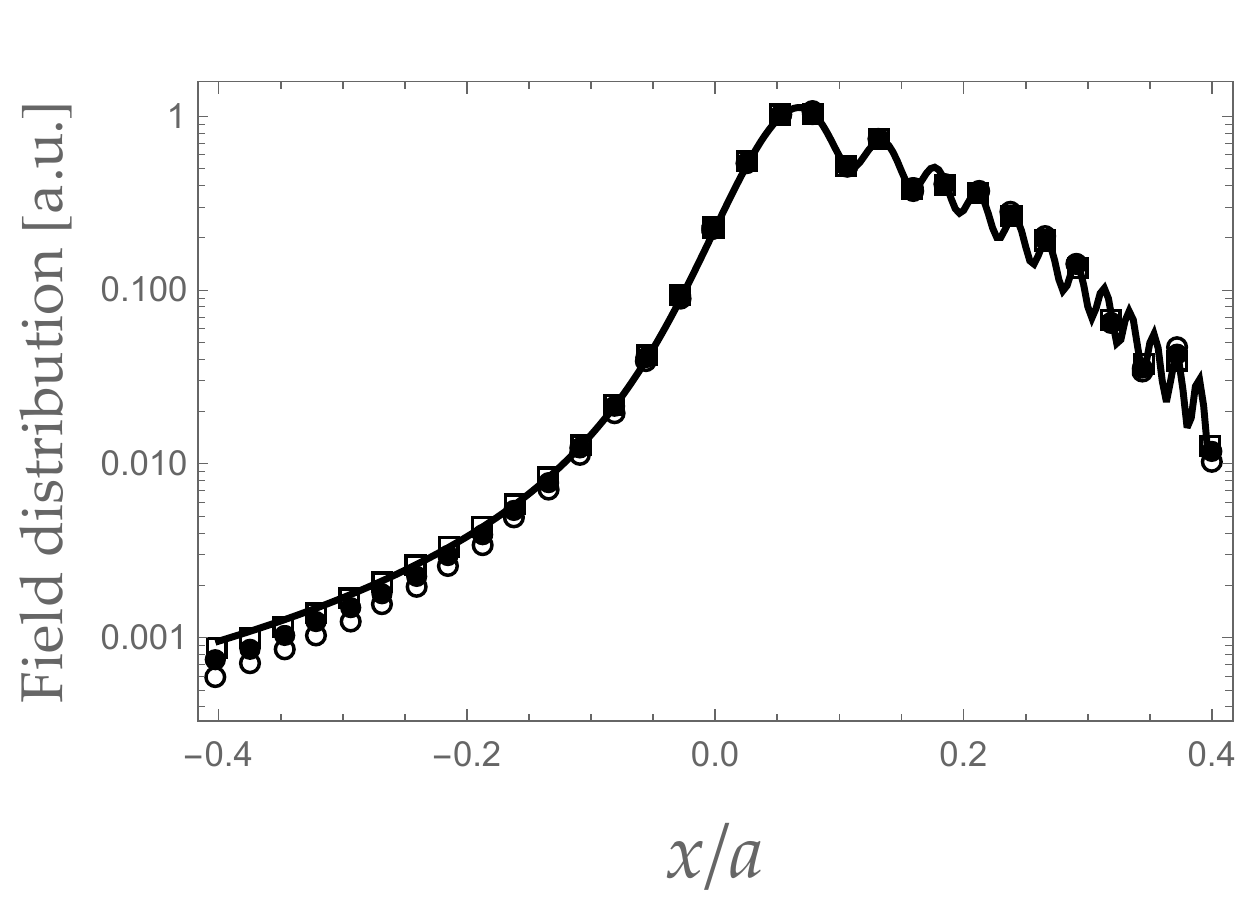}}
\caption{Behaviour of the optical intensity distribution $|\psi_G+\psi_{\rm BDW}|^2$ as a function of 
the normalized variable $x/a$ for $\alpha=5$ (open circles), 10 (dots), and 50 (open squares), together with the exact analytical expression
for the infinite edge found in~\cite{Pearson/McGill/Kurtin/Yariv/1969}, for the experimental values of the parameters given in the second column
of Tab.~\ref{Tab:table.1}..}
\label{Fig:YarivSperimentale.2}
\end{figure}

\begin{figure}[!ht]
\centerline{\includegraphics[width=8cm,angle=-0]{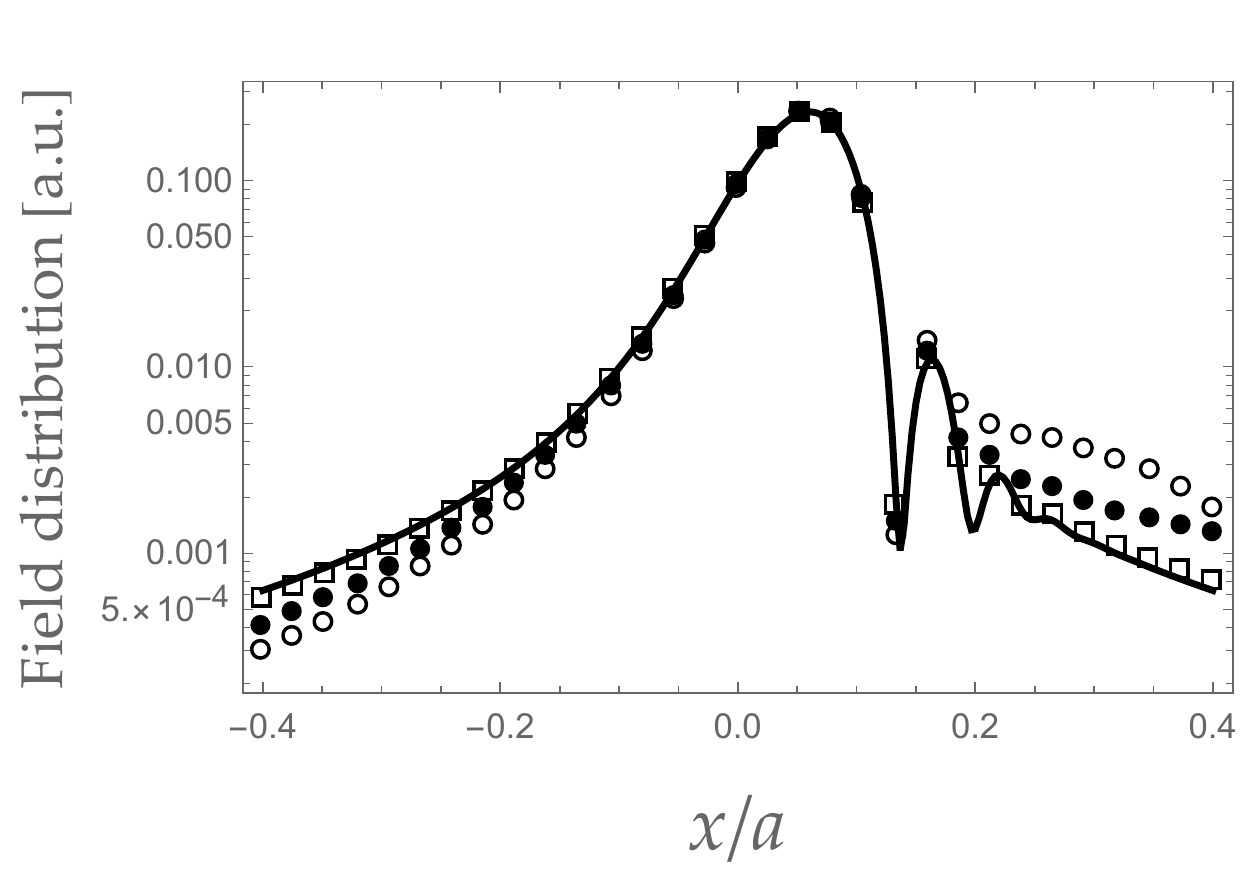}}
\caption{The same as in Fig.~\ref{Fig:YarivSperimentale.2} but for the experimental values of the parameters given in the third column
of Tab.~\ref{Tab:table.1}.}
\label{Fig:YarivSperimentale.3}
\end{figure}

In Fig.~\ref{Fig:YarivSperimentale.2} the optical intensity distribution $|\psi_G+\psi_{\rm BDW}|^2$ is plotted as a function of 
the normalized variable $x/a$ for $a/w_0=5$ (open circles), 10 (dots), and 50 (open squares), together with the exact analytical expression
for the infinite edge found in~\cite{Pearson/McGill/Kurtin/Yariv/1969}, for the experimental values of the parameters given in the second column
of Tab.~\ref{Tab:table.1}. The same has been done in Fig.~\ref{Fig:YarivSperimentale.3}, but for the experimental values of the parameters given in the third column
of Tab.~\ref{Tab:table.1}. In doing the above figures, the integrals in Eqs.~(\ref{Eq:FresnelPropagatorConvolution.3.New.4}) and~(\ref{Eq:FresnelPropagatorConvolution.3.New.5})
have been numerically evaluated simply on parametrizing the aperture as $(1+\cos t,\sin t)$, $t\in [0,2\pi]$.

\begin{figure}[!ht]
\centering
\begin{minipage}[t]{8.4cm}
\includegraphics[width=4.1cm,angle=-0]{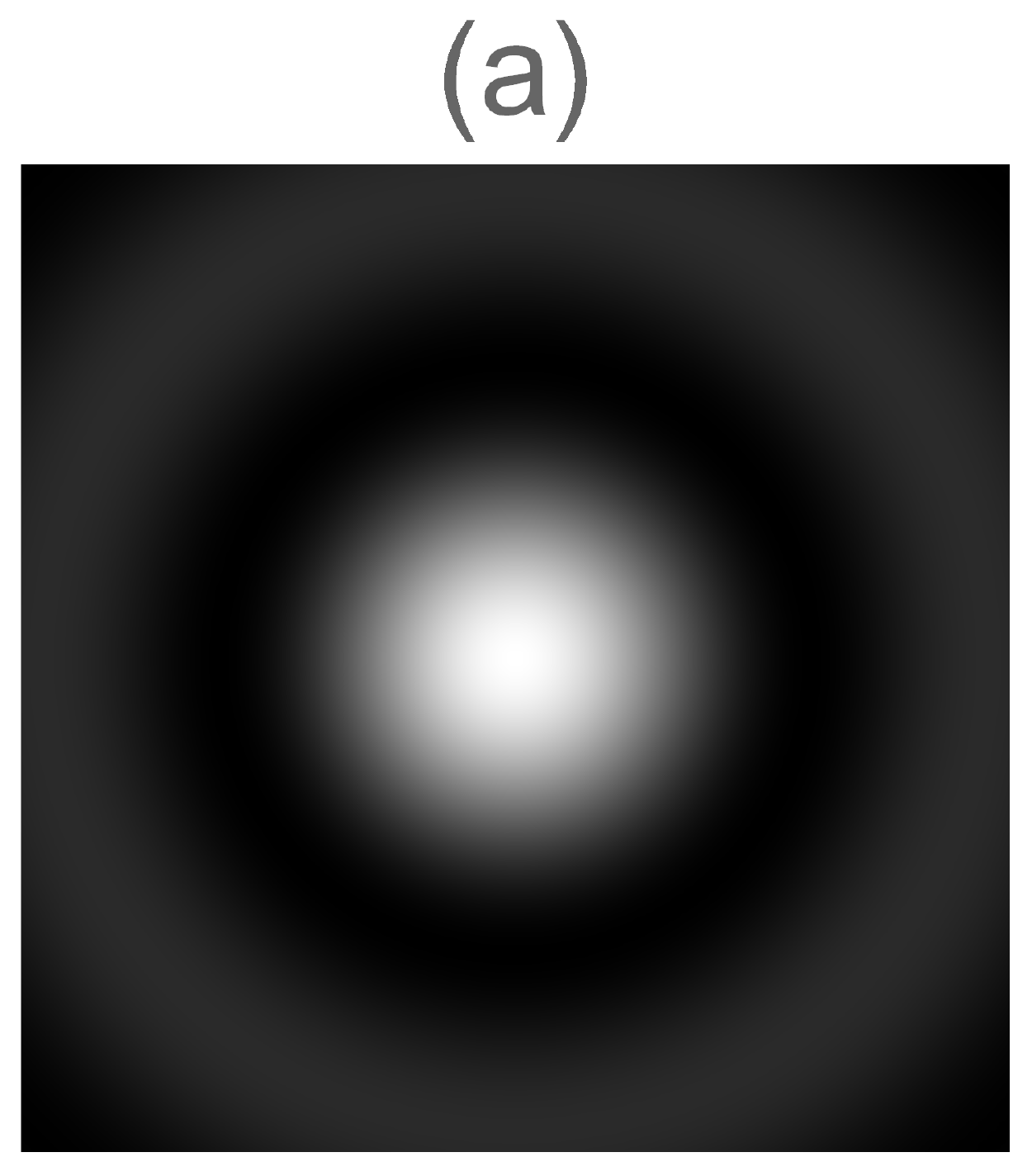}
\hfill
\includegraphics[width=4.1cm,angle=-0]{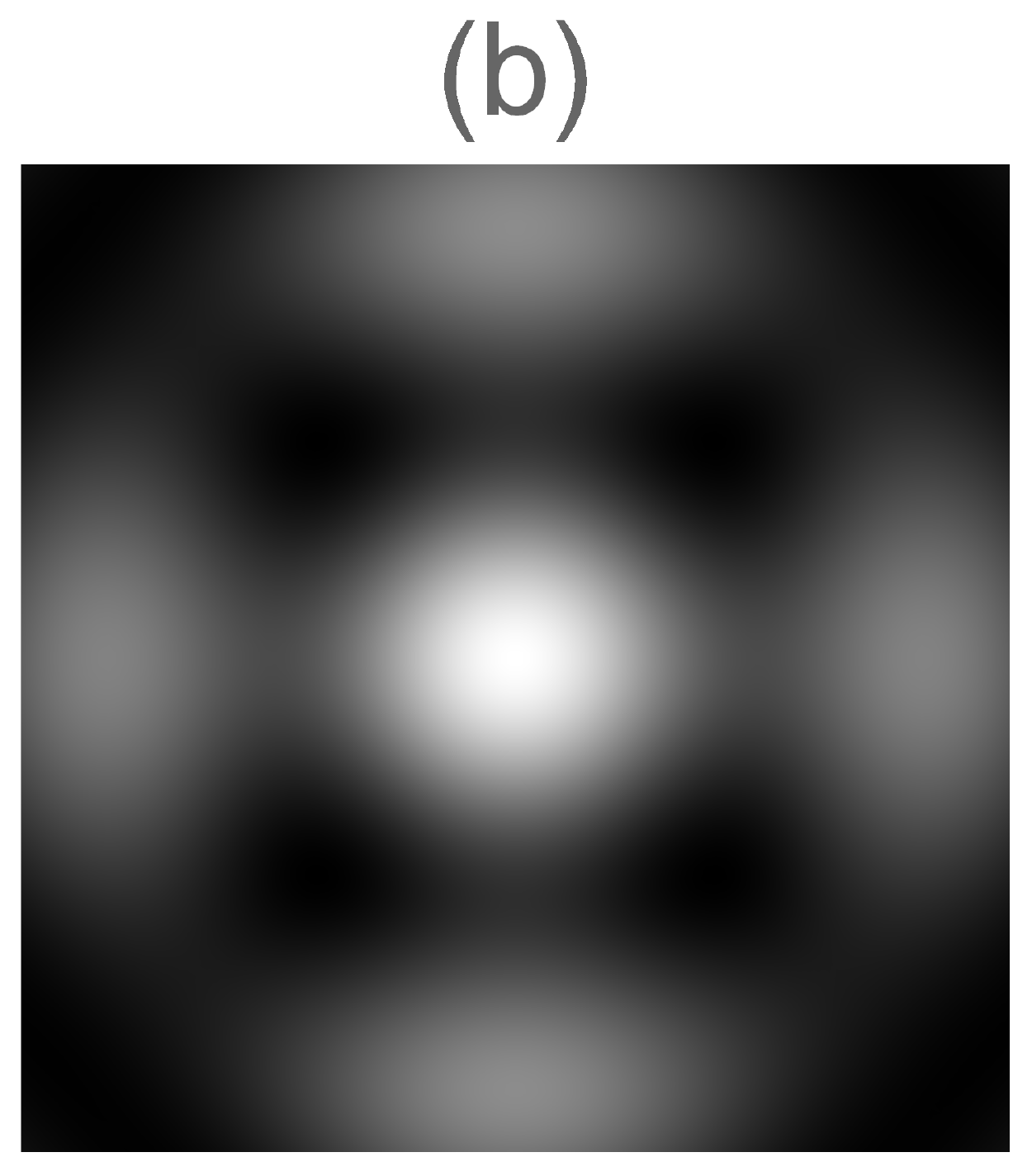}
\end{minipage}
\begin{minipage}[t]{8.4cm}
\includegraphics[width=4.1cm,angle=-0]{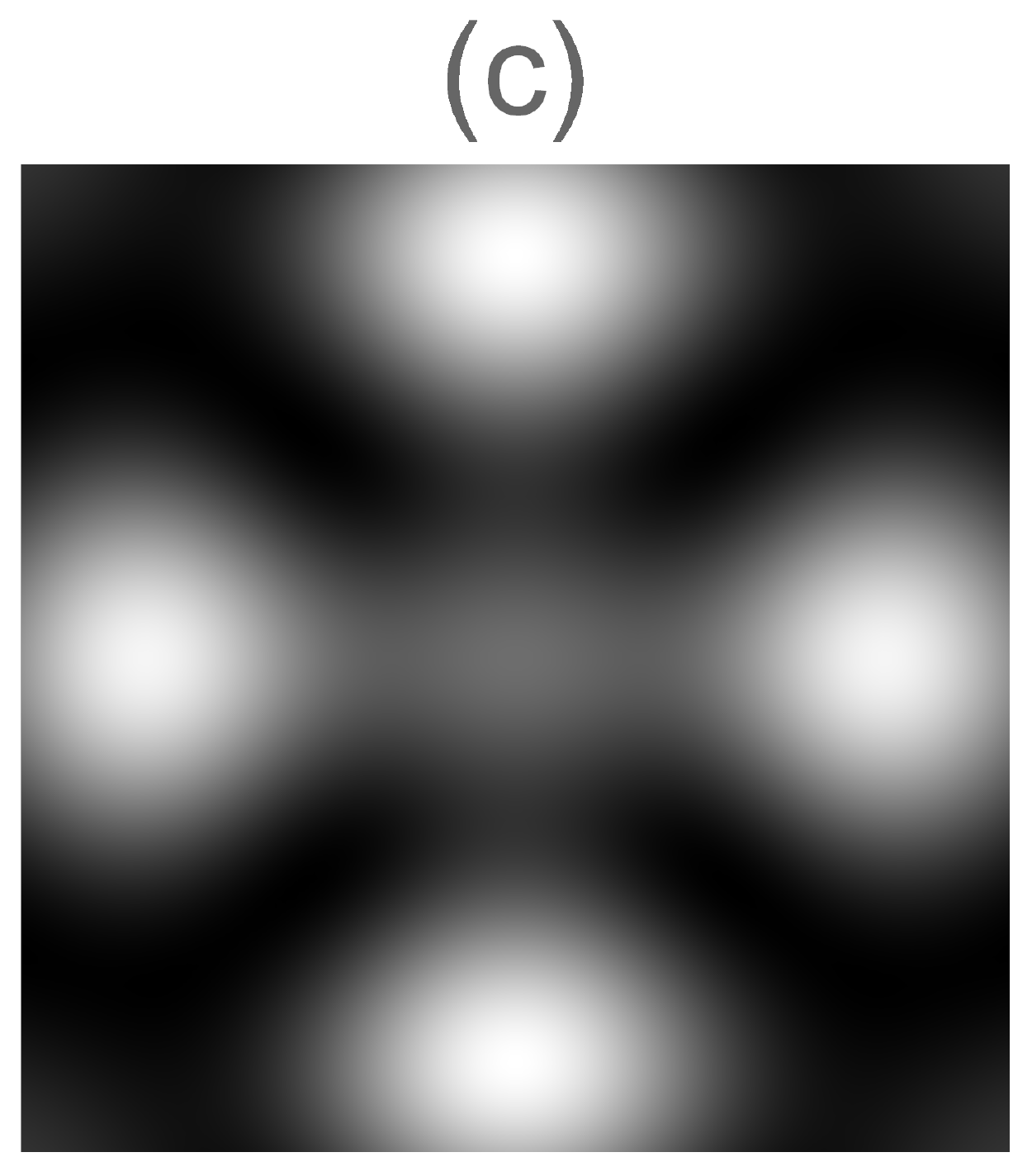}
\hfill
\includegraphics[width=4.1cm,angle=-0]{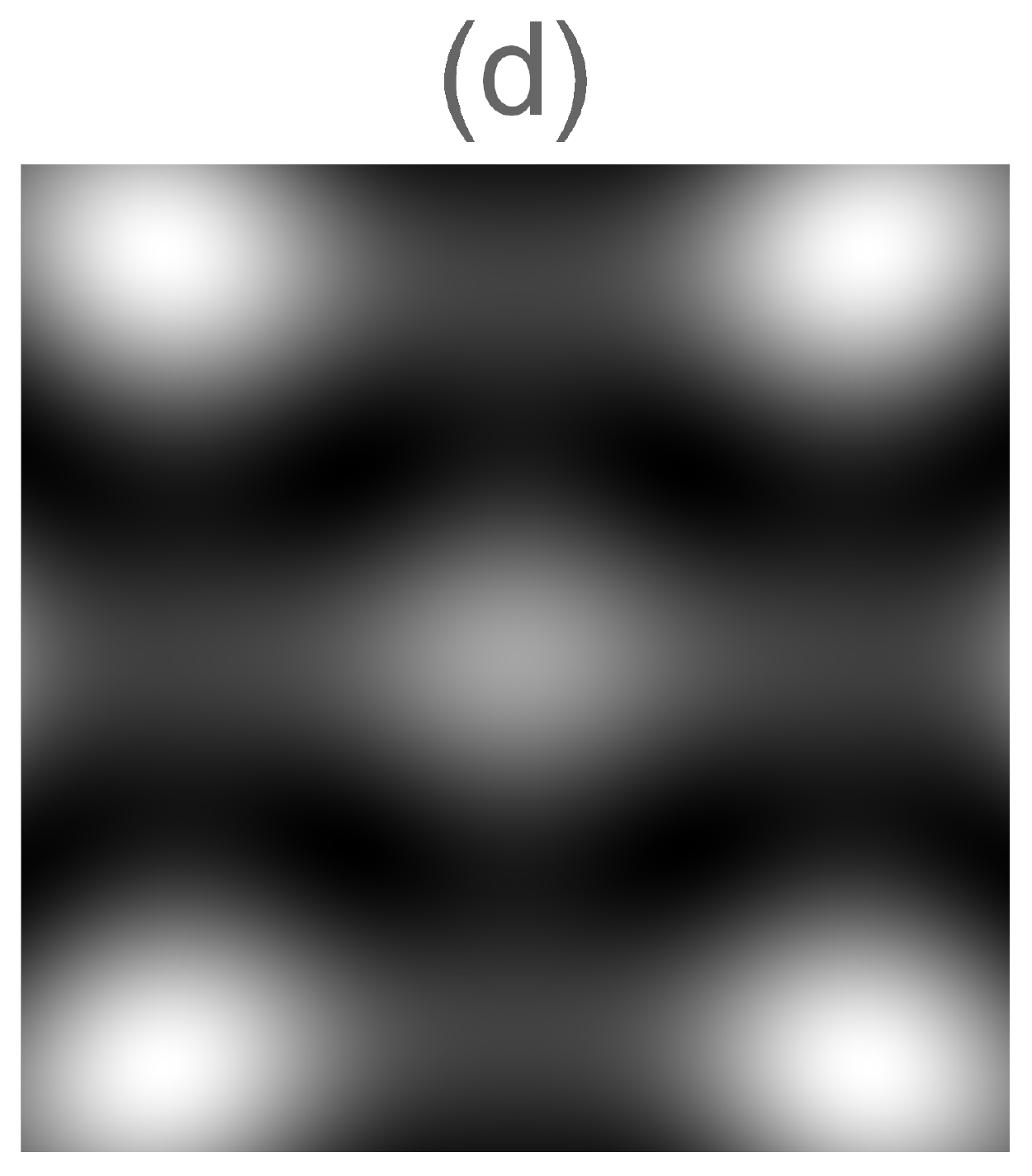}
\end{minipage}
\caption{Two-dimensional maps of the optical intensity of the diffracted wavefield produced,
close to the $z$-axis, at the normalized propagation distance $z=1/20$, by a tilted circular plate 
illuminated with a Gaussian beam whose waist plane is at the plate, $\alpha=1$ and (see Fig.~\ref{Fig:Misaligned})
$\theta=0$ (a), $\theta=\pi/10$ (b),  $\theta=\pi/8$ (c), and $\theta=\pi/6$ (d).}
\label{Fig:Focusing.1.2}
\end{figure}

As a second (and final) application, the paraxial theory here developed will be now implement in order to explore the finest details of the diffractive patterns 
produced, within the geometrical shadow, by tilted  opaque circular plates illuminated by collimated  Gaussian beams. 
To this end consider the situation depicted in Fig.~\ref{Fig:Misaligned} where in place of the aperture we consider a circular plate tilted by a 
small angle $\theta$ and placed for simplicity on-axis with respect the mean propagation distance of a collimated Gaussian beam having
the spot size equal to the plate radius, i.e., such that $\alpha=1$, and whose waist plane coincides with the plate plane. 
We shall explore the diffraction patterns close to the beam axis $z$, where the geometrical wavefield $\psi_G$ is expected to be null. 
Let us start with a perfectly aligned circular plate. In the case of a plane-wave illumination, it is well known the diffracted pattern at any transverse plane
to display a central bright spot, the celebrated Arago (or Poisson) spot. However, when the impinging field is Gaussianly shaped, the axial spot
becomes darker and darker on increasing the propagation distance $z$. This can  be viewed on analytically evaluating the BDW field $\psi_{\rm BDW}$ on-axis 
through Eq.~(\ref{Eq:FresnelPropagatorConvolution.3.New.5}). Since the Gaussian waist is at the plate plane, i.e., 
$D=0$, Eq.~(\ref{Eq:FresnelPropagatorConvolution.3.New.5}) gives at once the following on-axis intensity distribution:
\begin{equation}
\label{Eq:GaussianEllipse.OnAxis}
\begin{array}{l}
\displaystyle
|\psi_{\rm BDW}|^2\,=\,\dfrac {\exp(-2\alpha^2)}{1+z^2}\,,
\end{array}
\end{equation}
which, of course, becomes unitary in the plane-wave limit $L\to \infty$.
According to catastrophe optics~\cite{Berry/Upstill/1980}, the bright axial spot represents a highly unstable field configuration
which is made possible only by the perfect axial symmetry of the system composed by the diffracting plate and the illuminating wavefield. 
Accordingly, even a weak perturbation of such symmetry would be sufficient to produce dramatical topological changes on the 
resulting diffractive patterns. 
%
%
To visually appreciate these changes induced by the axial symmetry breaking,
in Fig.~\ref{Fig:Focusing.1.2} two-dimensional maps of the intensity of the diffracted wavefield close 
to the $z$-axis  are shown for a tilted circular plate  placed at the waist plane ($D=0$) of a Gaussian beam having 
$\alpha=1$. The diffraction patters are generated at the normalized propagation distance $z=1/20$, for 
$\theta=0$ (a), $\theta=\pi/10$ (b),  $\theta=\pi/8$ (c), and $\theta=\pi/6$ (d).

\begin{figure}[!ht]
\centerline{\includegraphics[width=8cm,angle=-0]{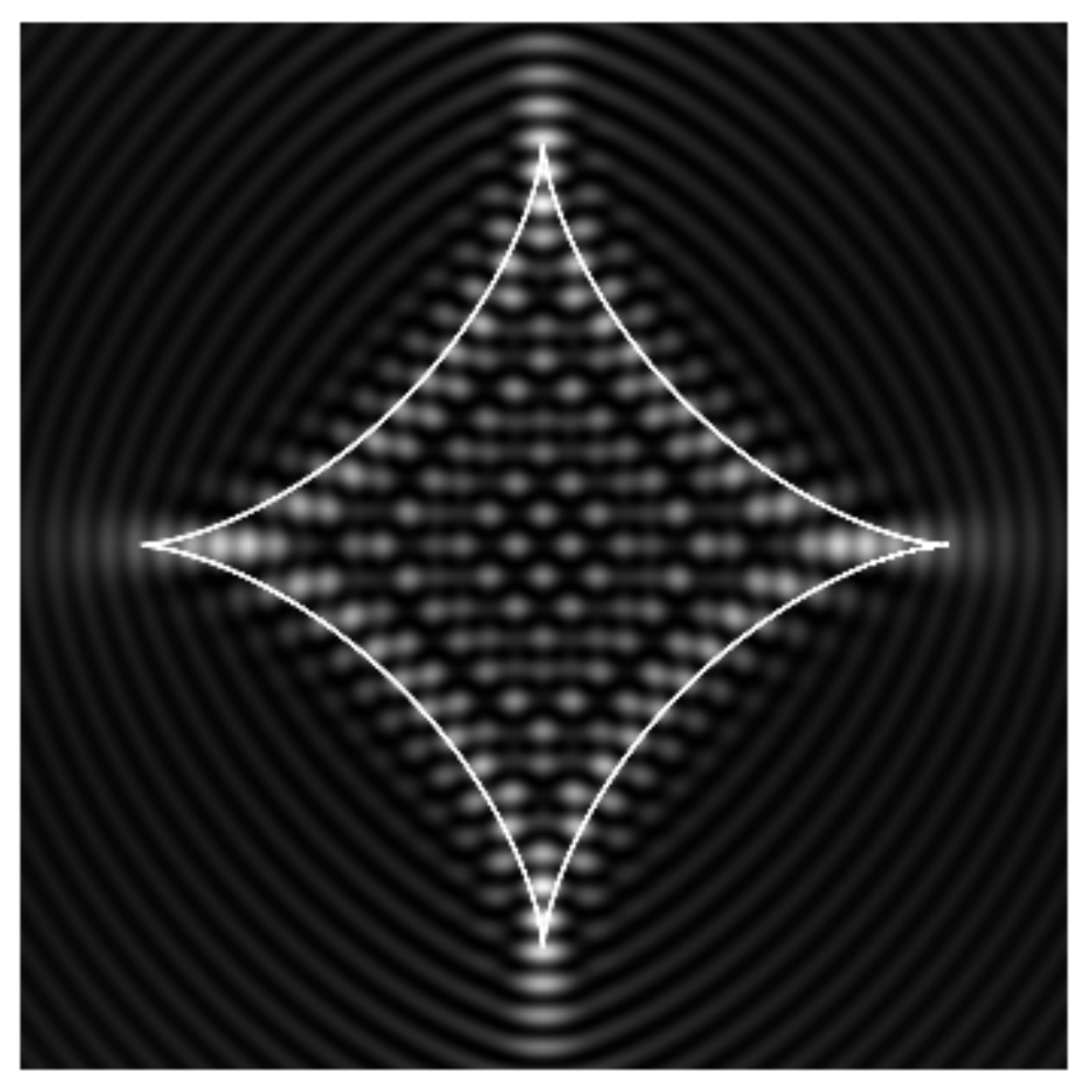}}
\caption{ Two-dimensional map of the transverse diffractive pattern generated for a 
tilting angle $\theta=\pi/6$ at $z=500/35333$ for a Gaussian beam with 
$\alpha=1$. The chosen value of $z$ corresponds to a Fresnel number $u$ given by $12\pi$. 
This has been done to allow a direct comparison with Fig.~10 of~\cite{Borghi/2014}, where the plane-wave diffraction 
by an identical elliptic plate was numerically investigated to reproduce the experimental 
results shown in Figs.~11 of~\cite{Coulson/Becknell/1922b}.
The white solid curve represents the cusp-shaped geometrical evolute of the elliptic boundary.}
\label{Fig:Focusing.1.3}
\end{figure}

Figure~\ref{Fig:Focusing.1.2}a is nothing but a blow-up of the Gaussian Poisson spot 
produced by the non tilted plate. For nonexperts of catastrophe optics it would be far from being trivial to 
appreciate the topological instability of the Poisson spot, which is led to ``explode'' into a stable, cusp-shaped 
configuration even by a small perturbation of the diffraction setup.
To this end, in Fig.~\ref{Fig:Focusing.1.3} the two-dimensional map of a bigger portion of the transverse diffractive 
pattern generated for the tilting angle $\theta=\pi/6$ is shown at $z=500/35333$ for a Gaussian beam with 
$\alpha=1$. The chosen value of $z$ corresponds to a Fresnel number $u$ given by $12\pi$. This has been done 
to allow a direct comparison with Fig.~10 of~\cite{Borghi/2014}, where the plane-wave diffraction by an identically elliptic
plate was numerically investigated to reproduce the experimental results shown in 
Figs.~11 of~\cite{Coulson/Becknell/1922b}. It is interesting to note how also for the more realistic Gaussian
illumination the BDW wavefield tends to focus onto the cusp-shaped geometrical evolute of the elliptic boundary
(the white solid curve), in agreement with the theoretical general prescriptions provided by catastrophe optics~\cite{Borghi/2016}

\section{Conclusions}
\label{Sec:Conclusions}

Diffraction theory is a milestone  of classical optics since more than two centuries. 
However, in the last few years  an unexpectedly renewed interest in new, still 
unexplored aspects of sharp-edge diffraction is grown~\cite{Weisman/Fu/Goncalves/Shemer/Zhou/Schleich/Arie/2017,Goncalves/Case/Arie/Schleich/2018,Narag/Hermosa/2018,Chen/Lee/Kang/Monro/Lancaster/2018,Ye/Xie/Wang/Feng/Sun/Zhang/2018,Borghi/2018}. The recently revisitation of paraxial sharp-edge diffraction 
developed in~\cite{Borghi/2015,Borghi/2016,Borghi/2017} has here been employed  to propose a paraxial version of 
Miyamoto-Wolf's  theory for exploring the  light diffraction produced by arbitrarily shaped planar apertures (or plates) 
under  Gaussian beam illumination.
On invoking the paraxial approximation from the 
beginning, the mathematical formulation considerably simplifies with respect
the theory originally developed in~\cite{Miyamoto/Wolf/1962}. 
In particular, both the geometrical and the BDW components of the total diffracted wavefield turn out to be expressed 
through one-dimensional complex integrals whose practical implementation and numerical evaluation
issues are reasonably independent of the aperture (or plate) shape. As a consequence, the
present approach could constitute an agile and effective general purpose computational platform 
to deal with a broad spectrum of different scenarios. Gaussian diffraction from 
highly nonsymmetric apertures as well as the prediction of  light behaviour 
by  image formation systems under realistic conditions of illumination, 
are only a couple of interesting applicative perspectives.

From a purely theoretical point of view, the analysis carried out on the geometrical shadows 
produced on  illuminating sharp-edge apertures by Gaussian beams seems to confute some 
important and definitive conclusions conjectured in the past~\cite{Otis/1974}. 
In particular, we have rigorously proved that the shape of the geometrical shadow turns out to be a perfect 
scaled replica of the diffracting aperture only for  circular holes. 
For differently shaped apertures it is no longer possible to predict the exact shape of the boundary,
even though some numerical experiments seem to confirm that the geometrical component of the diffracted 
wavefield  achieves binary values also under Gaussian illumination.
Presently we do not possess a rigorous and definitive conclusive word about such a very  
interesting and still open theoretical question.

\acknowledgments

I wish to thanks Dr. Javier Portilla and Dr. Sergio Barbero Briones for 
stimulating discussions about the subject.

\appendix

\section{Proof of Eq.~(22)}
\label{Sec:AppendixA}

For room reasons, we shall detail only the case $|\xi| <1$, leaving to the reader to deal with $|\xi|>1$.
Consider first the following integral:
\begin{equation}
\label{Eq:AppendixA.1}
\begin{array}{l}
\displaystyle
\mathcal{J}(\xi)\,=\,\dfrac 1\pi\,\int_0^\pi\,
\dfrac{\mathrm{d} t}{1\,-\,\xi\,\exp(\mathrm{i} t)}\,,
\end{array}
\end{equation}
which, after trivial algebra, can be recast as
\begin{equation}
\label{Eq:AppendixA.2}
\begin{array}{l}
\displaystyle
\mathcal{J}(\xi)\,=\,
\dfrac 1\pi\,\int_0^\pi\,
\mathrm{d} t\,\dfrac{1\,-\,\xi\,\cos t}{1\,+\,\xi^2\,-\,2\xi\,\cos t}\,\\
\\
\displaystyle
+\,\dfrac{\mathrm{i} \xi}{\pi}\,\int_0^\pi\,
\mathrm{d} t\,\dfrac{\sin t}{1\,+\,\xi^2\,-\,2\xi\,\cos t}\,,
\end{array}
\end{equation}
and where the second integral can be evaluated elementarly, so to have
\begin{equation}
\label{Eq:AppendixA.3}
\begin{array}{l}
\displaystyle
\mathcal{J}(\xi)\,=\,
\dfrac 1\pi\,\int_0^\pi\,
\mathrm{d} t\,\dfrac{1\,-\,\xi\,\cos t}{1\,+\,\xi^2\,-\,2\xi\,\cos t}\,
+\dfrac{\mathrm{i}}{\pi}\,\log\dfrac{1+\xi}{1-\xi}\,.
\end{array}
\end{equation}
Consider now Eq.~(\ref{Eq:AppendixA.1}) in which, due to the fact that $|\xi|<1$, the integrand can be expanded as a geometric series, 
\begin{equation}
\label{Eq:AppendixA.4}
\begin{array}{l}
\displaystyle
\dfrac{1}{1\,-\,\xi\,\exp(\mathrm{i} t)}\,=\,\sum_{k=0}^\infty\,\xi^k\,\exp(\mathrm{i} k t)\,.
\end{array}
\end{equation}
On substituting from Eq.~(\ref{Eq:AppendixA.4}) into Eq.~(\ref{Eq:AppendixA.1}) and after changing the series with the integral we have
\begin{equation}
\label{Eq:AppendixA.5}
\begin{array}{l}
\displaystyle
\mathcal{J}(\xi)\,=\,
\sum_{k=0}^\infty\,
\xi^k\,\dfrac 1\pi\,\int_0^\pi\,
{\mathrm{d} t}\,\exp(\mathrm{i} k t)\,,
\end{array}
\end{equation}
which, on taking into account that
\begin{equation}
\label{Eq:AppendixA.6}
\begin{array}{l}
\displaystyle
\dfrac 1\pi\,\int_0^\pi\,
{\mathrm{d} t}\,\exp(\mathrm{i} k t)\,=\,
\left\{
\begin{array}{lr}
1\,,& k=0\,,\\
\\
\dfrac{\mathrm{i}}{k\pi}\,[1-(-1)^k]\,,
& k \ne 0\,,
\end{array}
\right.
\end{array}
\end{equation}
eventually gives
\begin{equation}
\label{Eq:AppendixA.7}
\begin{array}{l}
\displaystyle
\mathcal{J}(\xi)\,=\,
1\,+\,
\dfrac{2\mathrm{i}}{\pi}
\sum_{k=0}^\infty\,
\dfrac {\xi^{2k+1}}{2k+1}\,=\,\\
\\
\,=\,1\,+\,\dfrac{\mathrm{i}}{\pi}\,\log\,\dfrac{1-\xi}{1+\xi}\,.
\end{array}
\end{equation}
On comparing Eqs.~(\ref{Eq:AppendixA.2}) and~(\ref{Eq:AppendixA.7}), it then follows at once
\begin{equation}
\label{Eq:AppendixA.8}
\begin{array}{l}
\displaystyle
\dfrac 1\pi\,\int_0^\pi\,
\mathrm{d} t\,\dfrac{1\,-\,\xi\,\cos t}{1\,+\,\xi^2\,-\,2\xi\,\cos t}\,=\,1\,, \qquad\qquad |\xi| < 1\,.
\end{array}
\end{equation}

To deal with the case $|\xi|>1$, it is sufficient to recast Eq.~(\ref{Eq:AppendixA.1}) as follows:
\begin{equation}
\label{Eq:AppendixA.9}
\begin{array}{l}
\displaystyle
\mathcal{J}(\xi)\,=\,
1\,-\,\dfrac 1\pi\,\int_0^\pi\,
\dfrac{\mathrm{d} t}{1\,-\,\xi^{-1}\,\exp(-\mathrm{i} t)}\,,
\end{array}
\end{equation}
and to apply again the above procedure. In this way it is not difficult to prove that 
\begin{equation}
\label{Eq:AppendixA.10}
\begin{array}{l}
\displaystyle
\dfrac 1\pi\,\int_0^\pi\,
\mathrm{d} t\,\dfrac{1\,-\,\xi\,\cos t}{1\,+\,\xi^2\,-\,2\xi\,\cos t}\,=\,0\,, \qquad\qquad |\xi| > 1\,,
\end{array}
\end{equation}
which, together with Eq.~(\ref{Eq:AppendixA.8}), completes the proof of Eq.~(\ref{Eq:GaussianCircle.1}).

\end{document}